\documentclass[%
reprint,
superscriptaddress,
amsmath,amssymb,
aps,
prx,
]{revtex4-1}

\usepackage{graphicx}  
\usepackage{dcolumn}   
\usepackage{bm}        
\usepackage{verbatim}   
\usepackage{epstopdf}
\usepackage{color}

\usepackage{textgreek}

\newcommand{\be}{\begin{equation}}
\newcommand{\ee}{\end{equation}}
\newcommand{\bea}{\begin{eqnarray}}
\newcommand{\eea}{\end{eqnarray}}
\newcommand{\bfac}{beta factor}
\newcommand{\bfm}{\ensuremath{\beta_{\rm m}}}
\newcommand{\bfs}{\ensuremath{\beta_{\rm s}}}

\newcommand{\Eq}[1]{Eq.\,(\ref{#1})}
\newcommand{\Fig}[1]{Fig.\,\ref{#1}}
\newcommand{\Sec}[1]{Sec.\,\ref{#1}}
\newcommand{\Onlinecite}[1]{Ref.\,[\onlinecite{#1}]} 
\newcommand{\App}[1]{appendix\,\ref{#1}}

\newcommand{\WGA}{\ensuremath{\rm WG_A}}
\newcommand{\WGB}{\ensuremath{\rm WG_B}}

\newcommand{\Cm}{\ensuremath{C_{\rm m}}}
\newcommand{\Cs}{\ensuremath{C_{\rm s}}}
\newcommand{\Dm}{\ensuremath{D_{\rm m}}}
\newcommand{\Ds}{\ensuremath{D_{\rm s}}}
\newcommand{\Gwg}{\ensuremath{\Gamma_{\rm wg}}}
\newcommand{\Gng}{\ensuremath{\Gamma_{\rm ng}}}
\newcommand{\Gnr}{\ensuremath{\Gamma_{\rm nr}}}

\newcommand{\Gc}{\ensuremath{\Gamma_{\rm c}}}
\newcommand{\Gnc}{\ensuremath{\Gamma_{\rm nc}}}

\newcommand{\EL}{\ensuremath{E_{\rm L}}}

\newcommand{\bk}{\ensuremath{\mathbf{k}}}

\newcommand{\xQD}{\ensuremath{x_{\rm QD}}}

\newcommand{\Lt}{\ensuremath{L_\text{t}}}
\newcommand{\Lb}{\ensuremath{L_\text{t}}}

\newcommand{\Lp}{\ensuremath{L_\text{p}}}

\newcommand{\zs}{\ensuremath{z_\text{s}}}
\newcommand{\xs}{\ensuremath{x_\text{s}}}
\newcommand{\ys}{\ensuremath{y_\text{s}}}

\newcommand{\Hzt}{\ensuremath{\tilde{H}_z}}
\newcommand{\br}{\ensuremath{\mathbf{r}}}
\newcommand{\bn}{\ensuremath{\mathbf{n}}}

\newcommand{\Aft}{\ensuremath{\alpha_{\rm FT}}}

\newcommand{\Fx}{\ensuremath{F_x}}
\newcommand{\Fy}{\ensuremath{F_y}}
\newcommand{\Fz}{\ensuremath{F_z}}
\newcommand{\Fb}{\ensuremath{F_{\rm b}}}
\newcommand{\Ff}{\ensuremath{F_{\rm f}}}
\newcommand{\Fl}{\ensuremath{F_{\rm l}}}

\newcommand{\Pin}{\ensuremath{P_{\rm i}}}
\newcommand{\Pic}{\ensuremath{P_{\rm ic}}}
\newcommand{\Pec}{\ensuremath{P_{\rm ec}}}
\newcommand{\Rc}{\ensuremath{R_{\rm c}}}
\newcommand{\etac}{\ensuremath{\eta_{\rm c}}}
\newcommand{\etas}{\ensuremath{\eta_{\rm s}}}

\newcommand{\LC}{\ensuremath{{\cal L}}}

\begin{document}

\title{99\% beta factor and directional coupling of quantum dots to fast light in photonic crystal waveguides determined by hyperspectral imaging}
\author{L. Scarpelli}
\affiliation{School of Physics and Astronomy, Cardiff University, The Parade, Cardiff CF24 3AA, United Kingdom}
\author{B. Lang}
\altaffiliation{Current address: School of Physics and Astronomy, University of Nottingham, UK}
\affiliation{Quantum Engineering Technology Labs, School of Physics and Department of Electrical\&Electronic Engineering, University of Bristol, Nanoscience and Quantum Information Building, Tyndall Avenue, Bristol, BS8 1FD, United Kingdom}
\author{F. Masia}
\author{D. M. Beggs}
\author{E. A. Muljarov}
\affiliation{School of Physics and Astronomy, Cardiff University, The Parade, Cardiff CF24 3AA, United Kingdom}
\author{A. B. Young}
\author{R. Oulton}
\affiliation{Quantum Engineering Technology Labs, School of Physics and Department of Electrical\&Electronic Engineering, University of Bristol, Nanoscience and Quantum Information Building, Tyndall Avenue, Bristol, BS8 1FD, United Kingdom}
\author{M. Kamp}
\author{S. H\"ofling}
\author{C. Schneider}
\affiliation{Technische Physik, Physikalisches Institut and Wilhelm Conrad R\"ontgen Center for Complex Material Systems, Universit\"at W\"urzburg, Am Hubland, 97474 W\"urzburg, Germany}
\author{W. Langbein}
\affiliation{School of Physics and Astronomy, Cardiff University, The Parade, Cardiff CF24 3AA, United Kingdom}

\date{\today}
\begin{abstract}
Spontaneous emission from excitonic transitions in InAs/GaAs quantum dots embedded in photonic crystal waveguides at 5\,K into non-guided and guided modes is determined by direct hyperspectral imaging. This enables measurement of the absolute coupling efficiency into the guided modes, the \bfac, directly, without assumptions on decay rates used previously. Notably, we found \bfac s above 90\% over a wide spectral range of 40\,meV in the fast light regime, reaching a maximum of ($99 \pm 1$)\%. We measure the directional emission of the circularly polarized transitions in a magnetic field into counter-propagating guided modes, to deduce the mode circularity at the quantum dot sites. We find that points of high directionality, up to 97\%, correlate with a reduced \bfac\, consistent with their positions away from the mode field antinode. By comparison with calibrated finite-difference time-domain simulations, we use the emission energy, mode circularity and \bfac\ to estimate the quantum dot position inside the photonic crystal waveguide unit cell. 
 \end{abstract}

\maketitle
\section{\label{sec:intro} Introduction}
Quantum dots (QDs) embedded in photonic crystal waveguides (PCWGs) are a promising system to implement quantum technologies. Due to the broadband coupling\,\cite{MangaRaoPRB07,LecampPRL07}, the system can be used for high-efficiency on-chip single photon sources. Furthermore, the strong lateral confinement of light results in a significant longitudinal component of the electromagnetic mode field, which allows for local circular polarization, and therefore selective coupling of circularly polarized dipoles into a single mode \,\cite{Rodriguez-FortunoS13,JungePRL13,MitschNC14,SollnerNNA15,YoungPRL15,LodahlN17}. Recent experiments have shown QD spin-photon path conversion and photon path-dependent QD spin initialization using this mechanism\,\cite{ColesNC16,ColesPRB17}, which is robust against disorder\,\cite{LangPRA15}. In conjunction with the recently demonstrated spin-controlled photon switching\,\cite{JavadiNNa18} and super-radiant emission from two coupled QDs in a PCWG\,\cite{KimARX18}, these results show the potential of such a system for the implementation of scalable quantum technologies on chip\,\cite{LodahlROMP15,MahmoodianPRL16}. A fundamental requisite of quantum technology based on QD on PCWGs is that the spontaneous emission (SE) from the emitter couples exclusively to the designed channels of the system, which are typically waveguide (WG) modes, and not to other background channels creating losses. For a given WG mode, the probability of a QD exciton (QDE) to emit into the mode is called the \bfac, defined as
\be
\beta=\frac{\Gwg}{\Gwg+\Gng+\Gnr}
\label{Eq:BetaFactorDef}
\ee
where $\Gwg$ and $\Gng$ are the decay rates to the selected WG mode, and other, typically non-guided, modes, respectively, and $\Gnr$ is the non-radiative decay rate. In previous works, the \bfac\ was estimated using lifetime measurements \cite{Lund-HansenPRL08,ThyrrestrupAPL10,DewhurstAPL10,LauchtPRX12,HoangAPL12,ArcariPRL14}. The challenge in these measurements is to determine all three decay rates. A common approach assumed $\Gwg$ to be given by the difference between the decay rate of a QDE coupled to the WG mode, $\Gc$, and the decay rate of similar QDEs not coupled to the WG, $\Gnc$, so that $\Gwg=\Gc-\Gnc$, and  $\Gng+\Gnr=\Gnc$. This accuracy of this analysis depends on the accuracy of the underlying assumption that all QDEs have the same decay rate into non-guided modes, and the same non-radiative decay. The non-radiative decay \Gnr\ is dependent on QD charging and local defects and is thus determined by properties beyond the photonic environment. Assuming no influence of local defects or tunneling, it is negligible for neutral excitons in InAs/GaAs QDs. Even in charged excitons, where Auger processes provide a non-radiative decay, this rate in the order of $1\,\mu$s$^{-1}$ \cite{KurzmannNL16} is three orders of magnitude below typical radiative decay rates of 1\,ns$^{-1}$ \cite{LangbeinPRB04a}. The radiative decay into other modes \Gng, however, is likely to be significantly modified by the local dielectric environment of the QDs in PCWG structures, as recent calculations\,\cite{JavadiJOSAB18} have highlighted. Therefore, the  analysis reported in previous work is expected to exhibit significant systematic errors in the determined \bfac, as recently pointed out \cite{RigalAPL18}.

In the present work we use direct hyperspectral imaging to determine the emitted power, avoiding assumptions on decay rates altogether. The SE from QDs embedded in the PCWG along the WG is imaged onto the input slit of an imaging spectrometer, and is measured spatially and spectrally resolved. In this way, the SE guided by the WG and coupled at the ends of the WG by grating couplers into free space is measured together with the SE emerging from the QD directly into free space. The \bfac\ is determined using the relative emission powers, after correcting for the propagation losses and relative efficiencies of the couplers. Using an external magnetic field in Faraday (out of plane) direction, the QDE states are split into two spectrally resolvable transitions with opposite circular polarisation. Depending on the QD location within the unit cell of the PCWG, the two transitions couple differently to WG modes of opposite propagation direction. Using the spectrally resolved emission from the two couplers, the emission into the two counterpointing WG modes is measured. Using the powers of the two transitions emerging from the two couplers, we quantify the directionality \cite{SollnerNNA15, ColesNC16} of the emission, and deduce the WG mode directionality $D$ at the QD site. The statistical distribution of the determined \bfac\ versus $D$ over a large ensemble of QDs, which are expected to randomly sample the in-plane area of the PCWG, shows that $\beta$ above 90\% are mostly found for directionalities below 80\%, and vice versa. The experimental results are analysed using detailed electromagnetic simulations of the PCWG structure. Specifically, we calculate $\beta$ and $D$ versus position inside the PCWG, and comparing with experiments, we estimate the QD position within the PCWG unit cell on a 10\,nm length scale. This position is expected to affect the exciton dephasing, both via the interaction with surface states, and through a modification of the local phonon density of states \cite{TighineanuPRL18}.
 
\section{\label{sec:Methods}Sample and Methods}
\begin{figure*}
	\includegraphics[width=\textwidth]{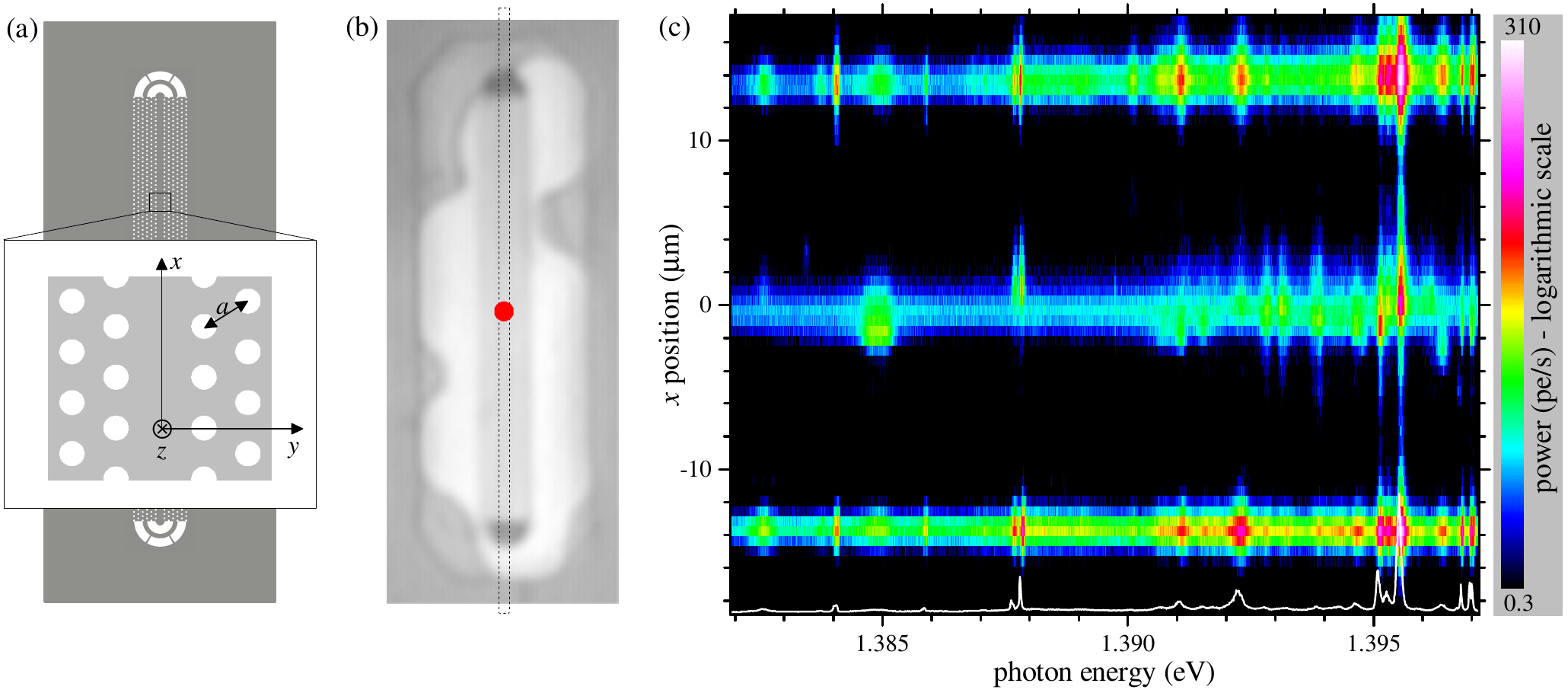}
	\caption{{\bf Sample and spectral imaging.} (a)\,Structure of the investigated PCWG. The zoom shows the lattice structure and the reference system used throughout the paper. (b) White light reflection image of $\WGA$. The red circle represents the spot of the excitation laser and the black dashed line indicates region corresponding to the input slit of the spectrometer. (c)\,Spectral image from the full CCD camera display, in units of photoelectrons per time and per pixel. The bottom and top regions are the QD emission into the WG mode and coupled to free space by the bottom and top couplers respectively. The central region is the direct free space emission from the QDs. White line: spectrum emitted from the bottom coupler.}
	\label{fig:Sample_SpectralImaging}
\end{figure*}
The investigated sample is a GaAs photonic crystal membrane of 125\,nm nominal thickness, with a single layer of InAs/GaAs QDs at the center. The QD area density was about $10^9$\,cm$^{-2}$, corresponding to an average distance of about 300\,nm, or about one QD per unit cell of the PCWG. The QDs are n-doped using a Si delta-doping layer with a density of about $10^{10}$\,cm$^{-2}$, 10\,nm below the QDs. Given the inhomogeneous size and spatial distribution of the QDs, this leads to a distribution of QD electron charging. Multiply charged QDs provide broader SE multiplets due to final state damping and spin-splitting. The emission lines analyzed were sharp lines which did not show a resolvable fine-structure splitting. Since typical values of fine-structure splittings for these QDs are in the few 10\,\textmu eV range \cite{LangbeinPRB04}, and considering the doping, we attribute these lines to negatively charged exciton transitions. Even though the charging is relevant for applications using the electron spin\,\cite{YoungPRL15,JavadiNNa18}, it is not important for the experimental results shown in this work. A fine-structure splitting much smaller than the magnetic-field induced Zeeman splitting of around 50\,\textmu eV, however, is important for the purity of the circular polarization of the transitions in a magnetic field \cite{BayerPRB02a}.

The PCWG is created by a line of missing holes in the periodic hexagonal pattern of round air holes of separation $a$, giving rise to guided modes within the two-dimensional photonic band gap for modes with dominating in-plane electric fields\,\cite{JohnsonPRB99}. An illustration of the investigated PCWGs is shown in \Fig{fig:Sample_SpectralImaging}a. The PCWGs investigated here are 100$a$ long, with $a=260$\,nm, and have 6 rows of holes on each side. The PCWGs are terminated by Bragg reflector couplers, which couple light propagating in the PCWG to free space\,\cite{FaraonOE08}. Two PCWGs (called $\WGA$ and $\WGB$) have been analyzed, which differ by their hole radius to period ratio $r/a$, being 0.24 and 0.26, respectively.  

The sample is mounted in a low vibration closed cycle cryostat (Montana Cryostation) on a XYZ piezo stage (Attocube) with a spatial resolution around 0.1\,\textmu m, allowing focusing and lateral alignment. For the measurements shown in this work, the sample temperature was 5\,K. A microscope objective (MO) with a numerical aperture NA of 0.85 is mounted inside the cryostat on the cold shield, having a temperature of about 30\,K. This avoids sample heating by thermal radiation, which can be an issue, specifically with membrane structures due to their reduced thermal contact to the substrate, when using the more common geometry with a room temperature window close to the sample and an external MO. A permanent magnet can be mounted below the sample to provide a magnetic field of $B_z$=0.45\,T, where $z$ labels the direction normal to the sample plane, as indicated in \Fig{fig:Sample_SpectralImaging}a. 

The QD SE was excited by a laser at a wavelength of $\lambda=633$\,nm focused onto the sample to a sub-\textmu m spot by the MO (see also \App{Sec:CarrierDiffusionLength}). The SE is filtered with a colour filter (Schott RG680) transmitting wavelengths above 680\,nm.
The SE from the QDs can be visualized by two imaging cameras, one for the near field (NF), imaging the real space at the sample, and one for the far field (FF), imaging the reciprocal space at the sample. For spectral imaging, the real space is imaged onto the input slit of an imaging spectrometer with a focal length of 1.9\,m, a 1200\,l/mm holographic grating of ($120 \times 140$)\,mm$^2$ size, 900\,nm blaze wavelength, and detected by a CCD (Roper Pixis) of $1340 \times 100$ square pixels of 20\,$\mu$m size. For all the measurements performed in this work, the input slit aperture was 20\,$\mu$m, corresponding to 639\,nm at the sample plane. The corresponding spectral resolution (full width at half maximum (FWHM)) is 8\,\textmu eV at 880\,nm (1.41\,eV). A white light K\"{o}hler illumination is integrated with the main PL setup, to simultaneously visualize the PCWG sample and the PL emission from the QDs. The origin of the $x$-axis along the WG is chosen at the center of the WG.

A reflection image of the sample is shown in \Fig{fig:Sample_SpectralImaging}b. The red spot indicates the excitation laser at a specific position along the WG. The excitation at a photon energy of 1.96\,eV (633\,nm wavelength) creates electron-hole pairs in the GaAs membrane, which subsequently relax by phonon emission towards the GaAs band gap around 1.52\,eV, before being captured into the  highly strained InGaAs wetting layer, where they further relax to the wetting layer band gap around 1.42\,eV, and finally into the QDs, which emit in the energy range between 1.37\,eV and 1.41\,eV. The direct imaging allows measurement of the diffusion length in the planar region of the sample, which was found to be about 3-4\,\textmu m on the membrane, as shown in \App{Sec:CarrierDiffusionLength}. The described carrier relaxation process is complex, with the formation of excitons in GaAs and the wetting layer also playing a role. Once captured into a QD, the carriers relax to the ground state within tens of picoseconds, from which they radiatively recombine, emitting a photon into the local photonic mode structure, consisting of the WG modes, confining light, and non-guided modes, rapidly escaping to free space on both sides of the slab. The emission is imaged onto the input slit of the spectrometer, indicated as black dashed line in \Fig{fig:Sample_SpectralImaging}b. Importantly, the WG has been aligned along the slit, in order to collect and image the emission from the whole WG and the couplers. The SE spectrum from the QDs is therefore detected spatially resolved along the WG, as exemplified in \Fig{fig:Sample_SpectralImaging}c, covering an energy range of about 15\,meV for a given spectrometer center position. Exciting at $x=0\,$\textmu m, we observe most of the free space emission close to the excitation, while signals around $x=\pm 13\,$\textmu m arise from the WG couplers, representing the QD SE into the WG modes. The spectrum from the bottom WG coupler is indicated as a white line, integrated over $y\in [-14.3, -11.4]$\,\textmu m. The center signal, close to the excitation position, is attributed to QD SE into the non-guided modes, excited by the carrier relaxation and diffusion processes described before. However, we also observed QD free-space emission from positions far away from the excitation spot, which we attribute to indirect excitation, where the wetting layer emission, which is coupled into the WG mode, is absorbed by QD excited states (see \App{sec:BetaVsExc}).
    
\section{\label{sec:BandStrucure}Photonic band structure}
\begin{figure*}
	\includegraphics[width=\textwidth]{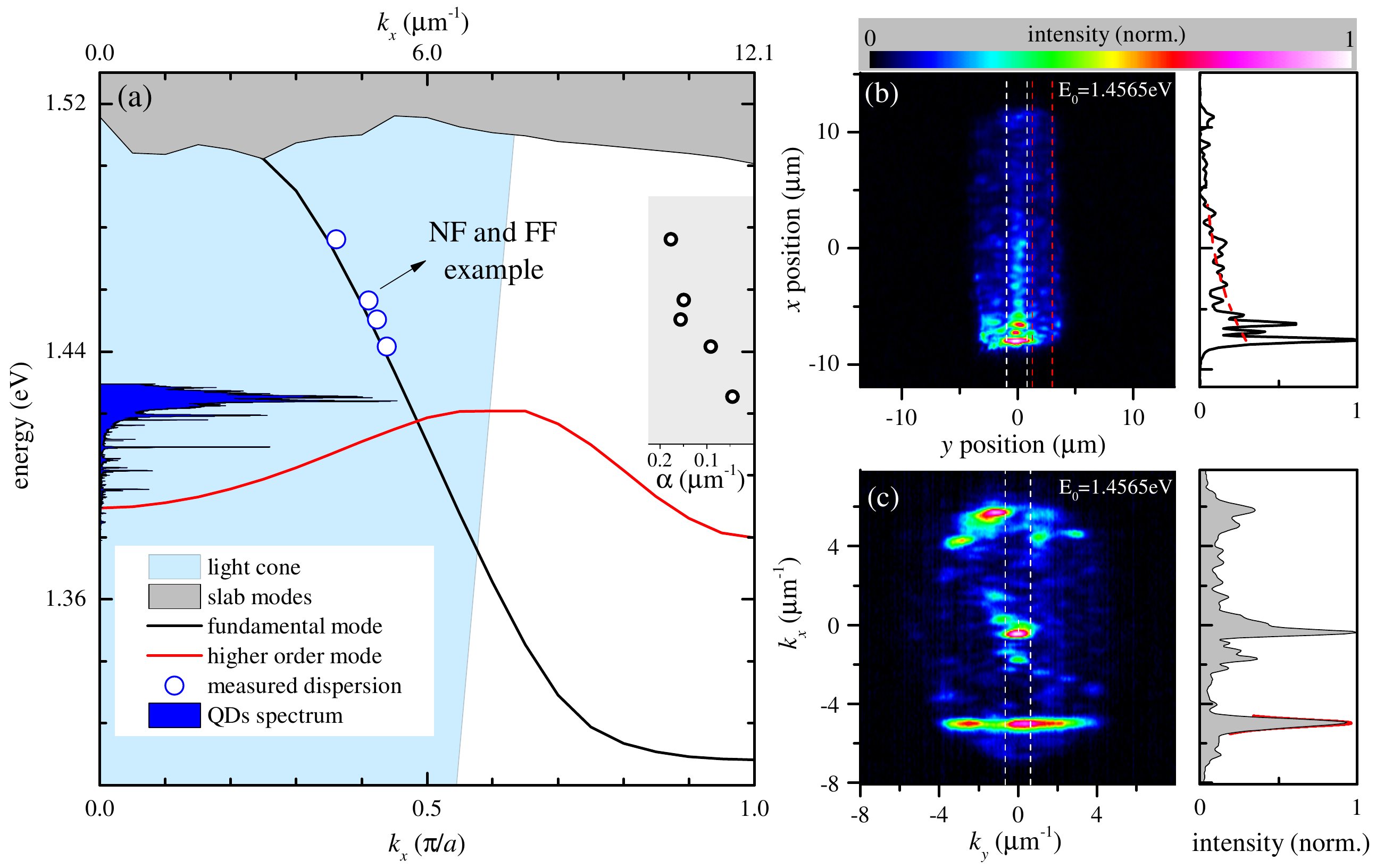}
	\caption{\label{fig:Dispersion} {\bf Photonic band structure and propagation imaging.} (a) Simulated photonic band structure of $\WGA$; dark-blue shaded area: QD emission spectrum; blue circles: measured dispersion from FF imaging; inset: WG loss $\alpha$ obtained from NF imaging. (b) NF image, on color scale as given, and the NF profile (right) obtained as the difference between a cut along the WG and a reference side cut (averages between the white dashed lines and red dashed lines in the image, respectively). Red dashed line: exponential fit. (c) FF image, and FF profile (right) obtained from a cut highlighted by white dashed lines in the image. Dashed red line: Lorentzian fit.}
\end{figure*}
Determining the photonic band structure of the investigated sample is crucial for a quantitative comparison with numerical simulations. In the literature, near-field scanning optical spectroscopy\,\cite{SukhorukovOE09} and interferometric techniques\,\cite{GersenPRL05} have been used to determine the guided mode dispersion in PCWGs. In other cases, the simulation parameters were adjusted to approximately reproduce the measured transmission window\,\cite{LauchtJAP12}. Alternatively, Fabry-P\'{e}rot fringes were used to calculate the group index of the guided mode\,\cite{ArcariPRL14}. 

Here, we use Fourier imaging \cite{LangbeinPRL99, LangbeinPRL02} to directly measure the band dispersion within the light cone\,\cite{LeThomasJOSAB07}. We use pulsed laser excitation, with center energies \EL\ corresponding to band wavevectors within the light cone, coupled to the WG via the bottom coupler. The laser polarization is set orthogonal to the WG direction, in order to select the fundamental mode (see \App{Sec:Grating Couplers}). To suppress the reflection of the excitation laser, we use a rectangular aperture in an intermediate image plane, corresponding to (8.7,19.0)\,\textmu m size at the PCWG. We measure both the NF and the FF of the emission along the WG as a function of energy, from which we can determine the propagation losses and the wavevector of the corresponding Bloch mode. An example of a NF measurement for \EL=1.4565\,eV is shown in \Fig{fig:Dispersion}b, with the corresponding FF measurement shown in \Fig{fig:Dispersion}c. The NF profile along the WG exhibits significant fluctuations, reflecting the coherent nature of the emission. After background subtraction, it can be fit with an exponential decay, as shown in the right panel of \Fig{fig:Dispersion}b, from which we obtain the WG loss $\alpha$, as further detailed in \App{Sec:NFanalysis}. The resulting $\alpha$ are given in the inset of \Fig{fig:Dispersion}a as function of \EL, showing an increase with \EL, which is attributed to both the radiative losses of the WG mode in the light cone and the absorption by the wetting layer, as discussed in more detail in \App{sec:LossesAnalysis}. 

Turning to the FF measurements, we note that the accessible range of the in-plane wavevector $\bk$ is limited by the NA of the MO to $|\bk|<k_0\text{NA}$, with the free-space wavevector $k_0=2\pi/\lambda$. The resulting cut-off is visible in \Fig{fig:Dispersion}c and from this we can calibrate the $k$-space of our measurements (see \App{Sec:FFCalibration}). The measured FF pattern shows a stripe at $k_x$ around $-5$\,\textmu m$^{-1}$, elongated in $k_y$ direction, which is the WG radiation loss. The WG mode wavevector is given by $k_x$, and the large extension in $k_y$ is due to the small extension of the WG mode in $y$. Knowing that the excited WG mode is propagating in positive $x$ direction, it is interesting to note that it exhibits a negative wavevector $k_x$ -- clear evidence of the negative group velocity of the WG mode, so that the phase velocity, given by $k_x$, is opposite to the group velocity, which is along the propagation direction.

The finite width in $k_x$ is due to the exponential decay of the field along the $x$ direction, the finite size of the imaged region, and the finite bandwidth of the excitation laser, as discussed in \App{Sec:FFResolution}. We determine the propagation wavevector by fitting the FF profile with a Lorentzian, as shown in the right panel of \Fig{fig:Dispersion}c. The FF profile is obtained by averaging over $k_y \in [-0.6,0.6]\,\mu \text{m}^{-1}$, indicated by the white dashed lines. The resulting $|k_x(\EL)|$ are shown in \Fig{fig:Dispersion}a, with the blue circles, noting that the reflection symmetry of the WG allows to use the absolute value.     

\begin{figure*}
	\includegraphics[width=\textwidth]{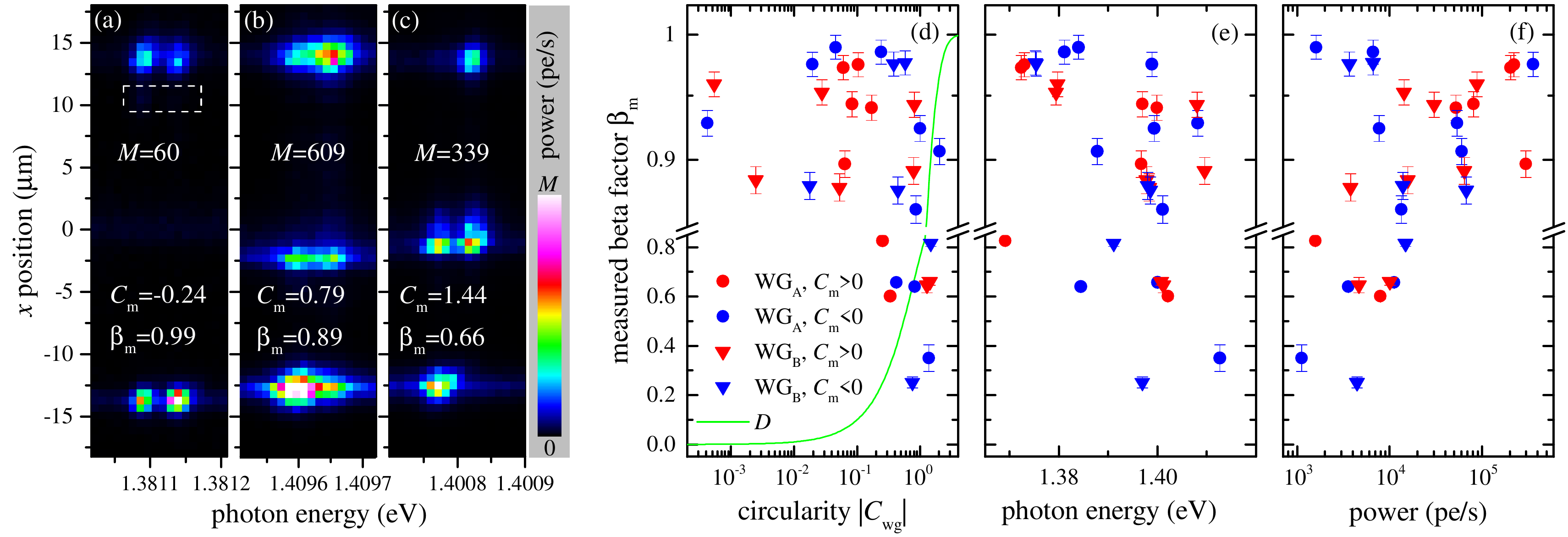}
	\caption{{\bf QD spectral images, \bfac, and directional emission.} (a),\,(b),\,(c)\,Spectral images of the SE from three different QDs located at sites of (a)\,low,\,(b)\,intermediate and (c)\,high WG mode circularity. Color scale as shown, from 0 to $M$. The corresponding $\Cm$ and $\bfm$ are given. The weak free space emission of the QD in (a) is highlighted by the dashed rectangle. (d),\,(e),\,(f)\,$\bfm$ as a function of (d) mode circularity $|\Cm|$, (e) QD peak energy, and (f) SE power for local excitation. In (d), the directionality $|\Dm|$ is shown (green line) as a function of $|\Cm|$.}
	\label{fig:SpectralImages}
\end{figure*}
We now use this measurement of the WG mode wavevectors to fine-tune the sample parameters used in simulations. The membrane thickness and hole diameters have been measured by scanning electron microscopy (SEM), and the photonic crystal (PC) period has a negligible fabrication error. We note that the refractive index of GaAs has previously been used as adjustable parameter, taking low temperature values of $\sqrt{12}$\,\cite{MangaRaoPRB07}, 3.5\,\cite{LauchtPRX12}, and 3.45 \,\cite{ArcariPRL14}. However, the GaAs refractive index is well known \cite{GehrsitzJAP00} and somewhat larger than these values. We therefore use in our simulations the known refractive index at low temperatures, including its dispersion. The QD layer embedded in the slab corresponds to less than 1\% of its thickness and is made of a similar material as GaAs, so that we neglect its effect on the refractive index. 

It is known that fabrication by selective etching, and oxidation of GaAs over time \cite{LukesSS72, DeSalvoJECS96}, can remove a surface layer of GaAs. We therefore use an effective thickness $d$ of a removed surface layer as a parameter, to match the simulations to the measurements, as detailed in \App{mode_finding}. The calculated band structure for $d=8$\,nm is shown in \Fig{fig:Dispersion}a, together with a typical free space SE of the QDs embedded in the PCWG. The measured spectrum consists of sharp lines at energies below 1.41\,eV, and a broad emission at higher energies, which we attribute to the wetting layer. Sharp lines superimposed to the wetting layer emission are attributed to localized excitons. 

\section{\label{Sec:ExperimentalResults} Results and discussion}
Using the spectral imaging as shown in \Fig{fig:Sample_SpectralImaging}, we identify the SE of individual QDs from the top and bottom couplers and from the QD position into non-guided modes. The excitation position can be adjusted along the WG to maximize the QD emission. The measurements were performed at a low excitation power to avoid multiexciton emission.

Examples of the SE detected for three different QDs are shown in \Fig{fig:SpectralImages}a-c, over a small spectral range covering the splitting of $\Delta_z\sim 50$\,\textmu eV created by the magnetic field. As the exciton has spin projection $S_z=\pm 1$, the Zeeman splitting between these states is given by $\Delta_z=2\mu_{\rm B}g_{\rm X} B_z$, with $\mu_{\rm B}$ the Bohr magneton. Using $B_z=0.45$\,T, we calculate an exciton g-factor $g_{\rm X} = 0.96$, comparable to the value of 1.2 reported in\,\cite{ColesNC16}.
The \bfac\ can be calculated as the ratio between the power emitted from the couplers and the total emitted power as
\be
\bfm=1-\frac{\sum_{j}P_{\text{fs}}^j}{\sum_{j}(P_{\text{t}}^j+P_{\text{b}}^j+P_{\text{fs}}^j)}\,,
\label{Eq:beta_factor}
\ee
where $P_{\rm b}$, $P_{\rm t}$, $P_{\rm fs}$ are the detected SE powers of a single QD from the bottom and top couplers and at the QD site, respectively, corrected for losses and relative coupler efficiencies (see \App{sec:LossesAnalysis}), and $j\in\{+,-\}$ labels the helicity of the QD transition. We note that we have not corrected the values for the simulated collection efficiency of free-space and coupler emission discussed in \App{subsec:freespace}, which would increase the \bfac, since the free space emission is collected typically twice as efficiently, depending on the QD position. Furthermore we determine the QD transition energy and its position along the WG, \xQD, as detailed in \App{sec:AnalysisSpectralImages}.

For the QD SE shown in \Fig{fig:SpectralImages}a, we find $\bfm=0.99\pm 0.01$. The QD is located close to the top coupler, as highlighted in the figure by the white dashed rectangle. Interestingly, the QD is visible for excitation at $x=0$, which we attribute to reabsorption of the wetting layer emission propagating along the PCWG. We verified that $\bfm$ is independent of the excitation position (see \App{sec:BetaVsExc}), as expected given that $\beta$ is determined by the local photon density of states, which is independent of the QD excitation pathway. \Fig{fig:SpectralImages}b,c show the SE of QDs with $\bfm=0.89\pm 0.01$ and $\bfm=0.66\pm 0.01$, respectively. The directional emission is evident in the asymmetry of the power from top and bottom coupler. The degree of directionality depends on the WG mode circularity $C$ at the QD site, which we calculate here from the measured powers as 
\be \Cm=\frac{1}{4}{\rm ln}\left(\frac{P^+_{\text{t}}P^-_{\text{b}}}{P^-_{\text{t}}P^+_{\text{b}}}\right)\,,
\label{Eq:wg_circularity} \ee
which by using the ratio of power ratios is independent of the efficiencies of the couplers. $C$ is related to the  directionality $D$ used in \Onlinecite{ColesNC16} by $D=\tanh(C)$, making it equal for small circularities but avoiding saturation for high circularities (see green line in \Fig{fig:SpectralImages}d).
At a circularly polarized point (C-point), $C$ diverges, while at a linearly polarized point (L-point), $C$ is zero.
Residual reflections from the couplers might limit the maximum value of \Cm. From the maximum measured \Cm, we can deduce that these reflections are below 5\%, see \App{sec:GratingReflection}. 

Notably, the  \bfm\ of the QDs shown in \Fig{fig:SpectralImages}a-c decreases with increasing $|\Cm|$. To investigate this further, we  determined circularity and \bfac\ for many QDs, in \WGA\ and \WGB, as shown in \Fig{fig:SpectralImages}d. We find that low \bfm\ are more likely at higher circularity. This is consistent with the fact that C-points require both transversal and longitudinal fields, and thus do not occur at the mode field anti-node. We find positive and negative circularities (red and blue symbols, respectively), randomly distributed for both investigated PCWGs. In the same figure, we show that the corresponding directionality $D$ is above 0.9 for some of the QDs, while in \Onlinecite{ColesNC16} the reported $D$ are below 0.8. Furthermore, we find \bfm\ above 0.9 over an emission energy range of about 40\,meV (see \Fig{fig:SpectralImages}e). In particular, in $\WGA$ we find QDs with $\beta$ above 0.9 for energies between 1.372\,eV and 1.408\,eV. We note that in most of this spectral range both fundamental and the higher order mode are predicted to be present, see \Fig{fig:Dispersion}. The coupling of a QD to these two modes depends on its energy and its position within the PCWG unit cell, as discussed in \App{Sec:ModeSeparation}. However, no obvious feature of the presence of the two modes is visible in the dependence of the \bfac\ on energy (\Fig{fig:SpectralImages}e). Importantly, these results show that efficient coupling occurs even in region of low group index within the light cone. In \Fig{fig:SpectralImages}f we plot \bfm\ as a function of the total SE power from the QD for excitation at \xQD, using for all QDs an excitation power of 0.5\,\textmu W at the sample. We find \bfm\ above 0.90 over three order of magnitudes in SE power. This indicates that the emission power is not dictated by the \bfac, but is governed mostly by the local carrier capture dynamics, which depends on the disorder landscape in the wetting layer close to the QDs. The circularity and QD energy also do not show a correlation with the coarse QD position along the WG, as shown in \App{sec:E_CwgVsxQD}.

\begin{figure*}
	\includegraphics[width=\textwidth]{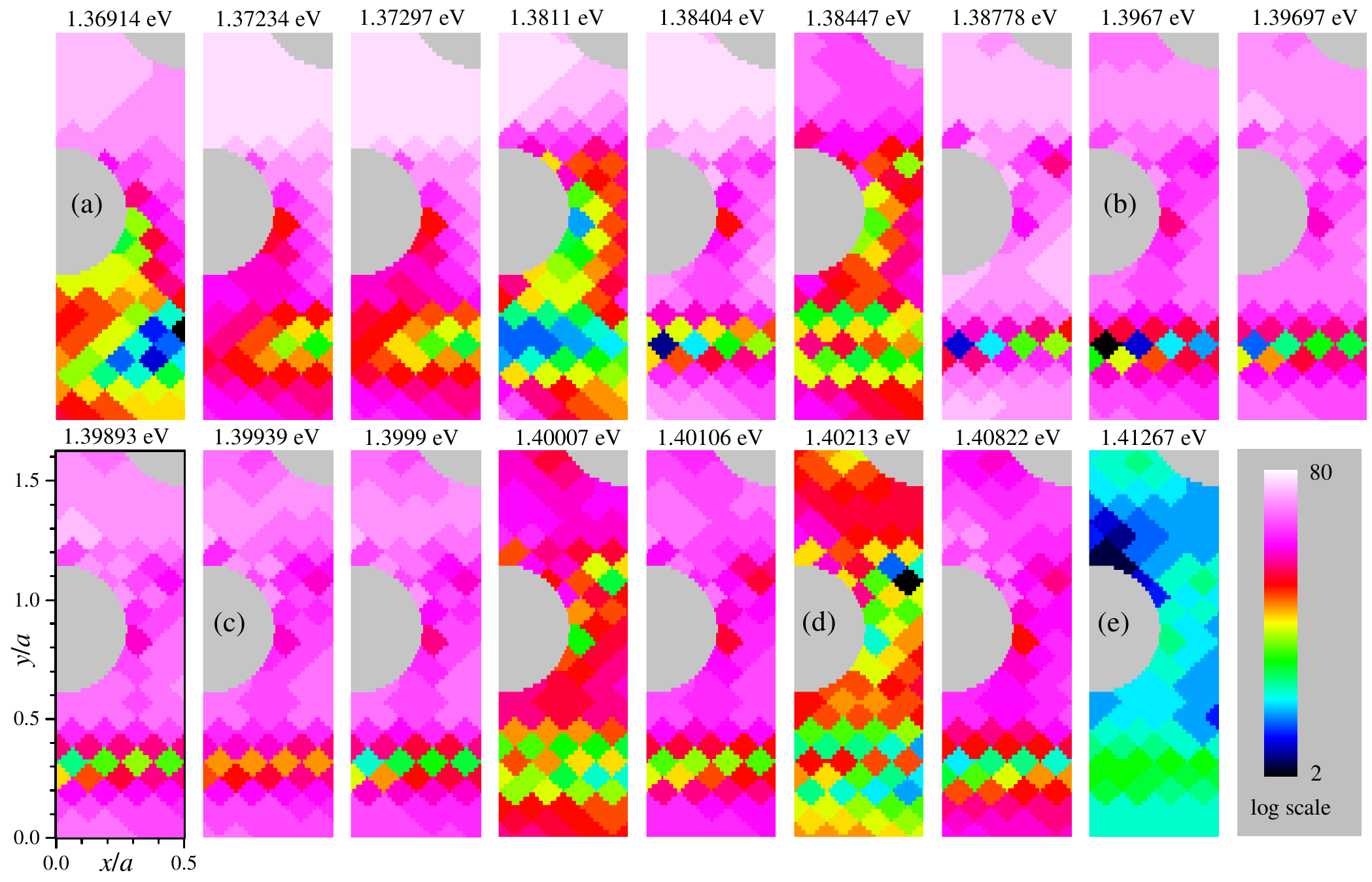}
	\caption{{\bf Determining the QD position in the PCWG unit cell.} Residual $\chi(x,y)$ of \bfac\ and circularity between simulations and experiment according to \Eq{Eq:ErrorPlot} for $\WGA$. Color scale as indicated. Positions $(x,y)$ inside the holes are indicated in gray. The values given are for the closest calculated point, in the center of the diamonds.}
	\label{fig:ErrorPlots}
\end{figure*}
The WG mode circularity at the QD site, together with the QD \bfac\ and energy, identifies possible positions of the QD inside the PCWG unit cell. We therefore performed finite-difference time-domain (FDTD) simulations for \WGA\ with $d=8$\,nm, extracting the simulated \bfac\ \bfs\ and circularity \Cs\ at positions across the PCWG unit cell, and at the energies of the measured QDEs, as detailed in \App{app:simulation}. The simulated positions are organised on a checkerboard grid with a step of $0.125a=32.5$\,nm. For a comparison with the measured quantities \bfm\ and \Cm, the simulated \bfac\ \bfs, and circularity \Cs, have been corrected to take into account the projections of the QD emission onto the fundamental and higher modes and the corresponding polarisation resolved collection efficiencies, which include the couplers, the NA of the MO, the input slit of the spectrometer and the efficiency of the spectrometer, as detailed in \App{app:simulation}, yielding $\tilde{\bfs}$ and $\tilde{\Cs}$. To identify the position of the QD, we compare simulations and experiment taking into account the experimental errors, by evaluating the normalized residual $\chi(x,y)$ for each QD, given by
\be
\chi(x,y)=\sqrt{\left(\chi_{\rm c}^2(x,y)+\chi_{\beta}^2(x,y)\right)/2},
\label{Eq:ErrorPlot}
\ee
with
\be
\chi_{\rm c}(x,y)=\frac{\Cm-\tilde{\Cs}(x,y)}{\Delta \Cm}, \quad \chi_{\beta}(x,y)=\frac{\bfm-\tilde{\bfs}(x,y)}{\Delta\bfm}.
\label{Eq:ErrorPlotTerms}
\ee
In these expressions, $\Delta\Cm$ and $\Delta\bfm$ are the experimental errors (standard deviation) for the measured $\Cm$ and $\bfm$, respectively. $x$ and $y$ are the simulated positions in the unit cell, according to the reference system given in \Fig{fig:Sample_SpectralImaging}a. Note that  agreement between simulation and experiment within the estimated errors is obtained for values of $\chi$ of the order of unity. In general, the minimum of the residual indicates the most likely QD position inside the unit cell. In \Fig{fig:ErrorPlots} the results are shown for QDs in $\WGA$, ordered with increasing energy. Generally, for most QDs we find positions with residuals below 5. Examples of QDs showing residuals below 2, labelled (a,b,d,e), enable to extract a well-defined position of the QD, with a precision of a few 10\,nm considering the region over which $\chi$ is increasing by unity (i.e. one standard deviation) from its minimum. These positions are distributed over the unit cell, as expected. One example in (c), shows $\chi>20$ for all simulated positions -- possibly, this QD is situated beyond the $y$ range covered in the simulations.       

\section{\label{sec:Conclusions}Conclusions}
In conclusion, we have shown that direct spectral imaging allows measurement of the \bfac\ and the directional emission, without assumptions on radiative decay rates, by using only the relative powers emitted into the WG mode and to free space. We found a maximum \bfac\ of $(99\pm 1)$\% in the fast light regime. Beta factors above 90\% are mainly found for quantum dots located at sites with small WG mode circularity, consistent with the fact that circular points occur away from the field antinodes. Using Fourier imaging to measure the band dispersion of the WG mode within the light cone, we calibrate FDTD simulations, allowing us to locate the QD positions inside the PCWG unit cell with a few 10\,nm precision, from their measured \bfac\ and circularity.
These results are promising for the suitability of the system for photon blockade and more advanced quantum technology, and the methods presented can be used to identify suited QDs. Furthermore, the position determination can be used to determine the precision of QD site-control techniques, important for the development of useful and scalable devices.
\section*{Acknowledgments}
This work was supported by the EPSRC under grant EP/M020479/1, and partially supported by the GW4 Accelerator "Southwest Quantum Dot Quantum Technologies GW4-AF3-004. The simulations were performed in the Advanced Computing Research Centre, University of Bristol. B.L. acknowledges support by an EPSRC studentship DTA 1407622. R.O. acknowledges her EPSRC fellowship  EP/N003381/1. Assistance in sample growth and fabrication by Monika Emmerlin, Johannes Beetz and Sebastian Maier is gratefully acknowledged.   

\section*{Contribution statement}
L.S. performed the measurements, analyzed the data, and drafted the manuscript, with support by F.M., W.L., and B.L.. B.L. performed the simulations, with support from D.B., E.A.M., L.S., and W.L.. A.B.Y., R.O.,  S.H., and C.S. provided the sample and participated in discussions and manuscript revision. W.L. guided the work.

\appendix 

\begin{figure*}
	\includegraphics[width=\textwidth]{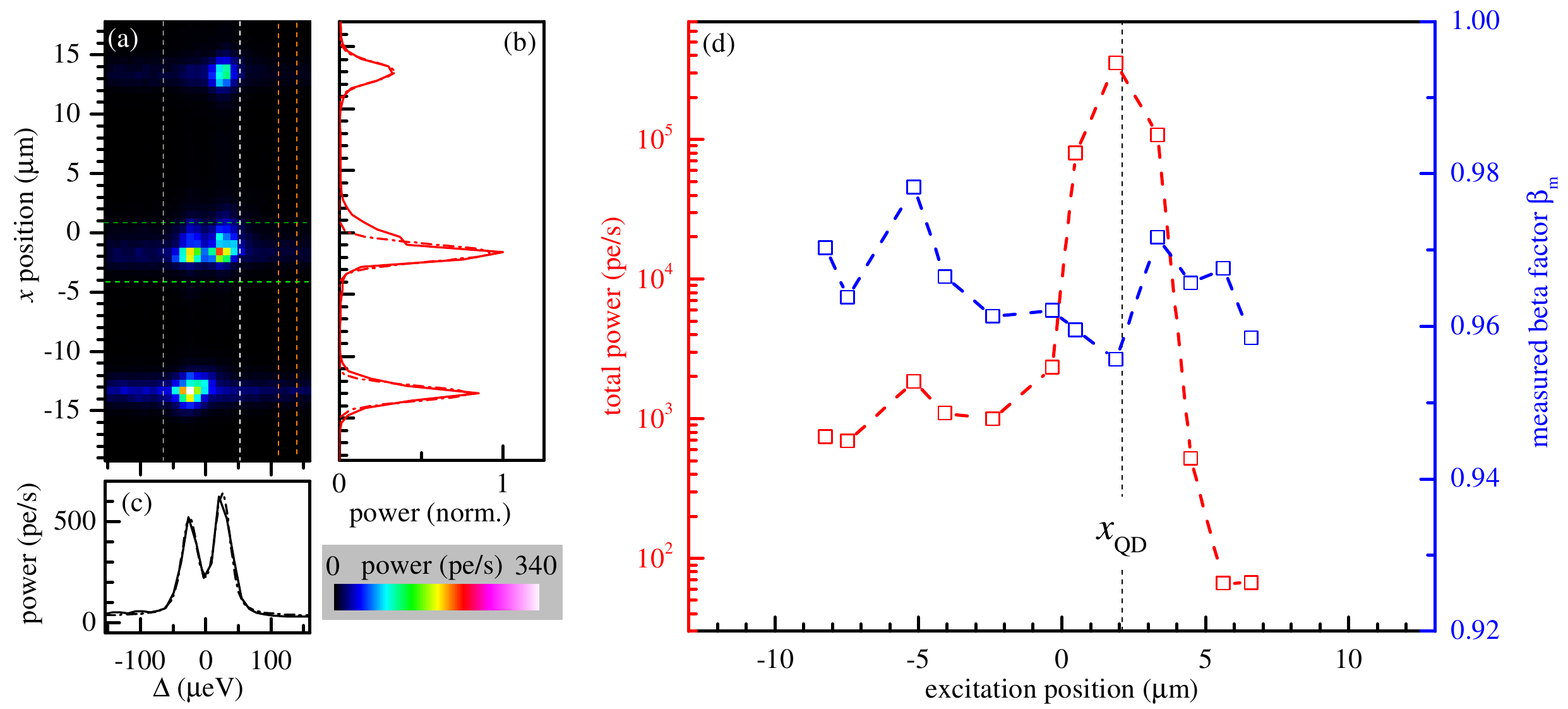}
	\caption{Analysis of spectral images. (a)\,Spectral image of QD shown in \Fig{fig:SpectralImages}c, centred 1.400796\,eV. (b)\,QD emission as a function of the position along the WG, taken as the spectral average around the QD emission energy (white dashed lines in (a)), and subtracting the background taken at a spectrally shifted region (orange dashed lines in (a)). (c)\,Spectrum of the QD free space emission (solid line), calculated as a spatial integral over the region given by the green dashed lines in (a); dash dotted lines are a fit to the data. (d)\,Total power and beta factor of a QD as a function of the excitation position along the WG. The black dashed line indicates the deduced QD position.} 
	\label{fig:AnalysisPlot}
\end{figure*}
\section{Analysis of spectral images}
\label{sec:AnalysisSpectralImages}
Here we discuss the analysis of spectral imaging data to retrieve the emission powers used in \Eq{Eq:beta_factor}  and \Eq{Eq:wg_circularity}. After subtracting the dark background, the data are divided by the integration time and multiplied by the CCD gain of 2 electrons per count, resulting in data as given in \Fig{fig:AnalysisPlot}a in photoelectrons per time and pixel. All data were taken in a magnetic field of $B_z=0.45$\,T, providing a Zeeman splitting into circularly polarized transitions, as visible in \Fig{fig:AnalysisPlot}a. In order to determine the total detected power emitted from the QD directly in free space, the  emission is integrated along the waveguide over the number of pixels contained in the region of interest. This region is highlighted in \Fig{fig:AnalysisPlot}a with dashed green lines. In order to find the peak positions and areas, we first fit the QD free space emission using a sum of two Voigt profiles, with equal linewidth parameters, as shown in \Fig{fig:AnalysisPlot}c. Using the obtained linewidths and energies, the emission from the couplers is fitted varying the amplitudes only. The areas of the fitted peaks, having units of photoelectrons per time, are taken as the measured powers $P_i^{j,\rm m}$,  with $i\in\{\text{b,\,fs,\,t}\}$ and $j\in\{+,-\}$. These powers are corrected for WG losses and relative coupler efficiencies, as described in \Sec{sec:LossesAnalysis}, to obtain the powers used in \Eq{Eq:beta_factor}. 

In order to obtain the QD position along the WG, \xQD, we first separate the QD signal from the excitation background by averaging over two regions, one including the QD signal and the other one spectrally shifted up by about 100\,\textmu eV containing the spectrally broad background from other emitters, as exemplified by the white and orange dashed lines in \Fig{fig:AnalysisPlot}a, respectively. Phonon-assisted transitions of the QD are spectrally broad and provide a background about two orders of magnitude below the zero-phonon line emission, which can be neglected in this analysis. Subtracting the excitation background, we isolate the QD signal, as shown in \Fig{fig:AnalysisPlot}b, which is then fitted using three Gaussian peaks, to determine the position of the two couplers and of the QD free space emission \xQD. The $x$ axis is calibrated using the known distance between the couplers. The position error resulting from the fit is typically around 100\,nm (standard deviation). Similarly, by fitting the background from other emitters, we determine the excitation position along the WG.

\section{\label{sec:BetaVsExc}\bfac\ versus excitation position}
In \Fig{fig:AnalysisPlot}d we show the total SE and the \bfac\ as a function of the excitation position along the PCWG for one specific QD. The total SE is the sum of the SE emitted from the bottom and top couplers and from the QD position into free space, as in \Fig{fig:SpectralImages}f. The total SE varies over three order of magnitude for different excitation positions along the WG, showing a maximum at the QD location \xQD\ determined from the free space SE as described in \App{sec:AnalysisSpectralImages}.  The \bfac\, determined versus exctation position, is also shown in \Fig{fig:AnalysisPlot}d, is constant within 1\% standard deviation. This shows that the determination of the \bfac\ is robust against changes of excitation conditions. This variation is taken as a lower limit for the error in the \bfac. We note that out of the 31 analyzed QDs, only a few show significant non-local excitation and all of these emit between 1.393\,eV and 1.377\,eV. Non-local excitation is attributed to re-absorption of wetting layer emission into a WG mode. We can speculate that the QD absorption occurs into the trion p-shell (see Supplementary Material of\,\Onlinecite{ColesPRB17} and \,\Onlinecite{BennyPRB12}), which is about 40\,meV above the s-shell, consistent with the energy separation between the wetting layer emission around 1.42\,eV and the QD emission. 

\section{\label{sec:E_CwgVsxQD}QD emission energy and WG mode circularity versus position}
\begin{figure}[]
	\includegraphics[width=\columnwidth]{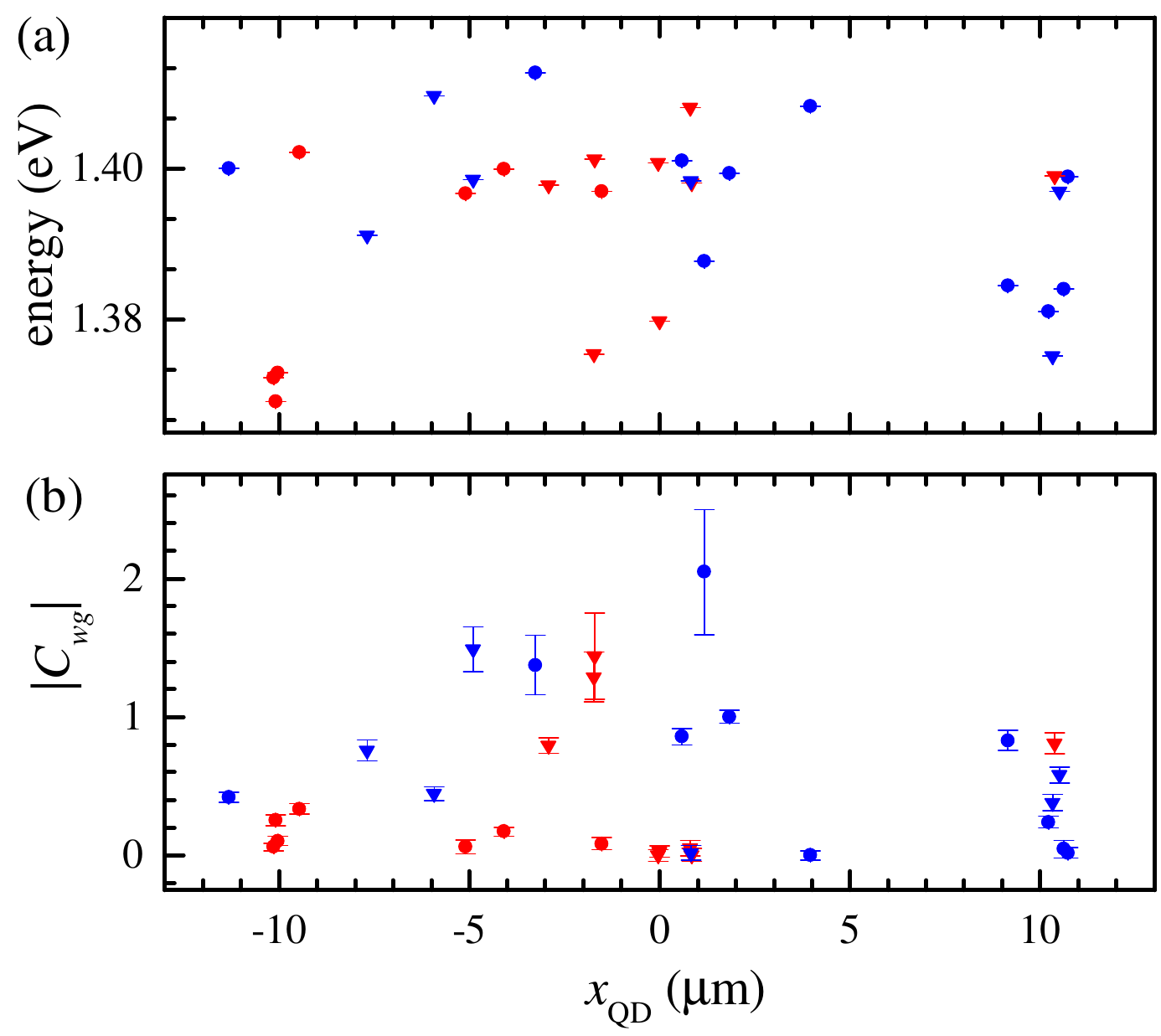}
	\caption{Spatial distribution of (a) QD energy and (b) circularity along the WG, versus QD position \xQD, determined as described in \App{sec:AnalysisSpectralImages}. Symbols as in \Fig{fig:ErrorPlots}d.} 
	\label{fig:QDxPositionPlots}
\end{figure}

One would expect that there is no correlation of energy and WG mode circularity with the QD position along the PCWG. The experimental data are given in \Fig{fig:QDxPositionPlots}, and exhibit a random distribution of energy and circularity along the QD position, consistent with this expectation.

\section{\label{sec:LossesAnalysis}Relative efficiencies of couplers}
\begin{figure}
	\includegraphics[width=\columnwidth]{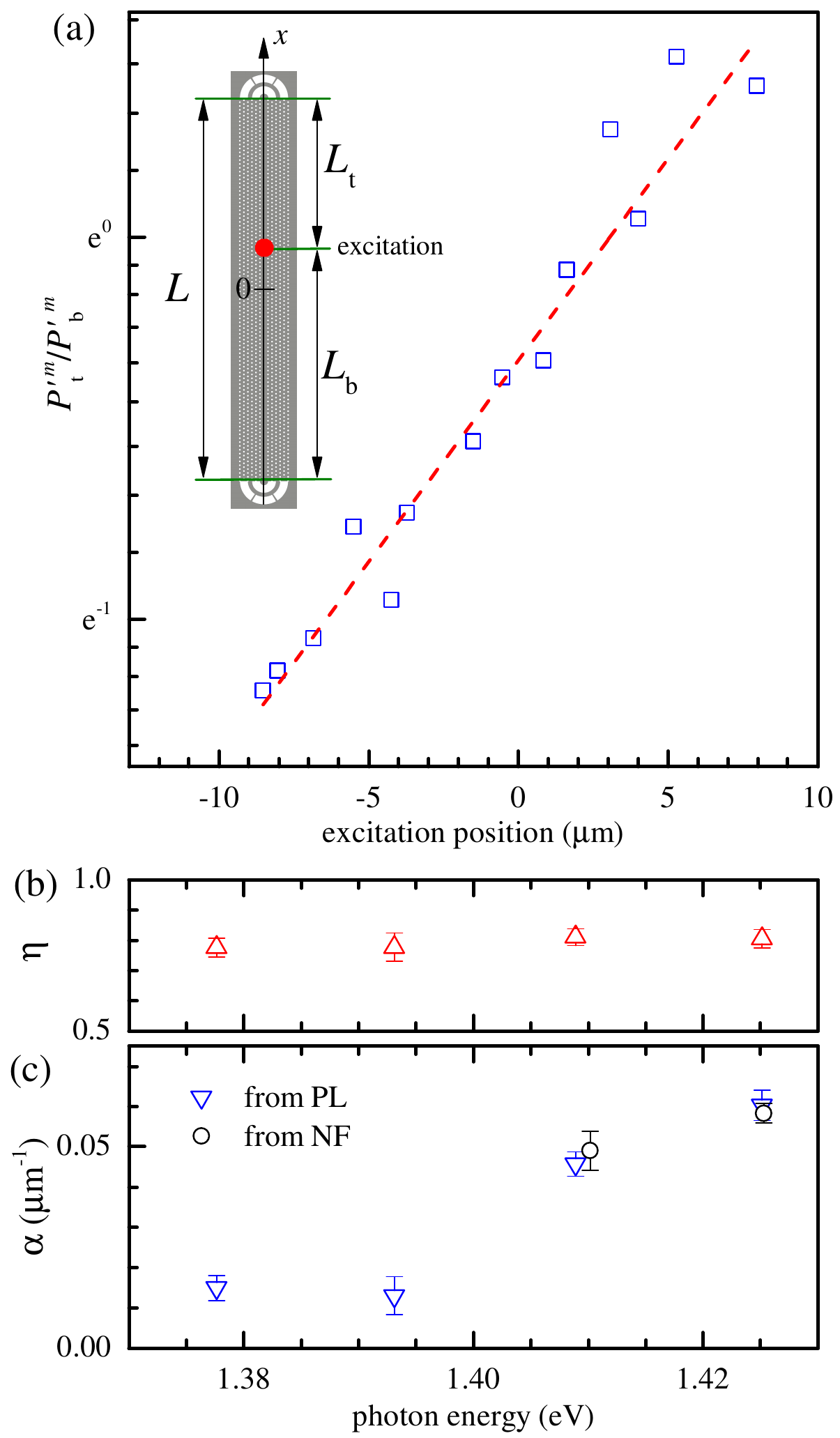}
	\caption{PCWG loss determined by SE measurements. Analysis described in \App{sec:LossesAnalysis} applied to \WGB. (a)\,Top to bottom emission ratio $P'^{\rm m}_{\rm t}/P'^{\rm m}_{\rm b}$ as a function of the excitation position for the spectral range centered at $\EL=1.425$\,eV. Inset: Sketch of geometry. (b)\,Relative collection efficiency from top to bottom coupler versus center energy. (c) Loss coefficient $\alpha$ versus center energy determined from the coupler emission ratio (triangles) and NF imaging (circles).} 
	\label{fig:LossesPlots}
\end{figure}
To determine the relative efficiencies of the two couplers, we perform the following analysis on data taken with magnetic field. For a given excitation position along the PCWG, we define the distances $\Lt=L/2-x$ and $\Lb=L/2+x$ between the excitation position $x$ and the top and bottom couplers, respectively, with the length $L=26\,\mu$m of the PCWG, as sketched in the inset of \Fig{fig:LossesPlots}a. Accounting for propagation losses $\alpha$, the SE powers emerging from the couplers are proportional to $\eta_{i}\exp(-\alpha L_{i})$, with the efficiency $\eta_{i}$ of the couplers, where $i\in\{\text{t},\text{b}\}$. Here we assume that most of the emission occurs at the excitation position, consistent with the experimental findings of the imaging along the WG, as discussed in \App{sec:BetaVsExc}. To apply the model, we spectrally integrate the SE from the bottom and top couplers over the range covered by the CCD camera ($\sim 15$\,meV), thereby averaging over the directionality of the individual QDs, resulting in the powers $P'^{m}_i$. The power ratio is fitted by
\be	
\frac{P'^{\rm m}_\text{t}}{P'^{\rm m}_\text{b}}=\eta e^{2\alpha x}
\label{LossesFunction}
\ee
with $\eta=\eta_\text{t}/\eta_\text{b}$, as shown in \Fig{fig:LossesPlots}a for the spectral range centered at $E_0$=1.425\,eV for $\WGA$. A good agreement between data and fit is observed. We repeated the analysis for the 4 spectral ranges considered in the experiment, and show in \Fig{fig:LossesPlots}b-c the resulting $\eta$ and $\alpha$ versus \EL. We find that $\eta$ has no significant dependence on \EL, and we use the average value of $\eta=0.79\pm 0.02$ to correct the emission intensities of all the data from this WG. The analysis has been repeated with $\WGB$, giving a similar result $\eta=0.62\pm 0.02$. The loss $\alpha$ increases with \EL, as expected from the increasing absorption by the wetting layer and the increasing radiation losses (see \App{loss}). Notably, the values are consistent with the loss measured by NF imaging, shown as black circles in \Fig{fig:LossesPlots}c.

The determined $\eta$ and $\alpha$ are used to calculate the powers $P_{\rm b}^j$ and $P_{\rm t}^j$ unaffected by loss and relative coupling efficiency, used in \Eq{Eq:beta_factor}, as
\be
\begin{aligned}
	P_{\rm b}^j=&P_{\rm b}^{j,\rm m} \exp\left[\alpha \left(\xQD+L/2\right)\right]\\
	P_{\rm t}^j=&\eta^{-1}P_{\rm t}^{j,\rm m}\exp\left[\alpha \left(L/2-\xQD\right)\right]
\end{aligned}
\label{Eq:GratingCorrection}	
\ee

\section{\label{sec:GratingReflection}Back-reflections at PCWG termination}
Back-reflections at the couplers can affect the observed circularity, since light emitted in a given direction is detected after the reflection at the opposite coupler. To estimate the influence of this effect for the investigated sample, we consider the measured powers of the QD with the highest mode circularity we found. To first order in the reflection coefficient $R$, the reflection of the measured power from the bottom coupler is adding to the measured power from top  coupler, and vice versa. The powers unaffected by the reflection are then given by $\tilde{P}_{\rm t}^{j} \approx P_{\rm t}^{j,{\rm m}}-RP_{\rm b}^{j,{\rm m}}$, and $\tilde{P}_{\rm b}^{j} \approx P_{\rm b}^{j,{\rm m}}-RP_{\rm t}^{j,{\rm m}}$. Requiring that these powers are positive, we find from the measured powers of all investigated QDs shown in \Fig{fig:SpectralImages} an upper limit for $R$ of about 3-5\,\%. The reflection leads to a systematic underestimation of the absolute circularity and directionality, and for 5\% is limiting the measured directionality to $|\Dm|<0.95$ and the measured circularity to $|\Cm|<1.8$. The dynamic range of our experimental data is sufficient to measure $|\Cm|$ above 4, and $|\Dm|$ above 99.9\%, and is thus not limiting the results presented.

\section{\label{Sec:NFanalysis}NF and FF analysis}
\begin{figure}[]
	\includegraphics[width=\columnwidth]{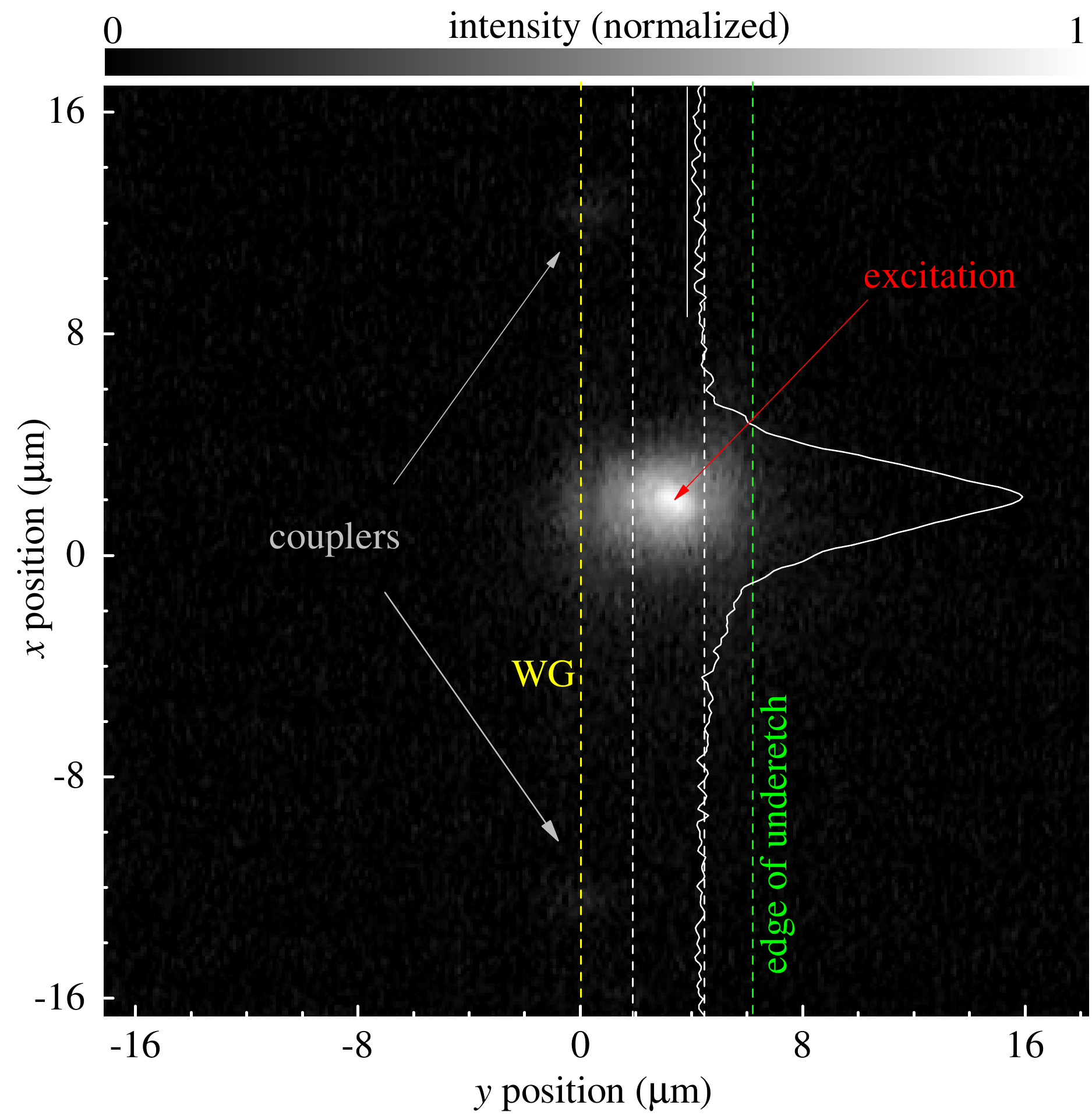}
	\caption{Carrier diffusion in the GaAs membrane. PL emission image for excitation on the PC membrane as indicated. Green dashed line: edge of underetched region. Carriers reaching the WG (yellow dashed line) excite the QDs and wetting layer, and emission is visible from the couplers. Solid white line: average intensity between white dashed lines, with zero given by baseline} 
	\label{fig:CarrierDiffusion}
\end{figure}
\begin{figure}[]
	\includegraphics[width=\columnwidth]{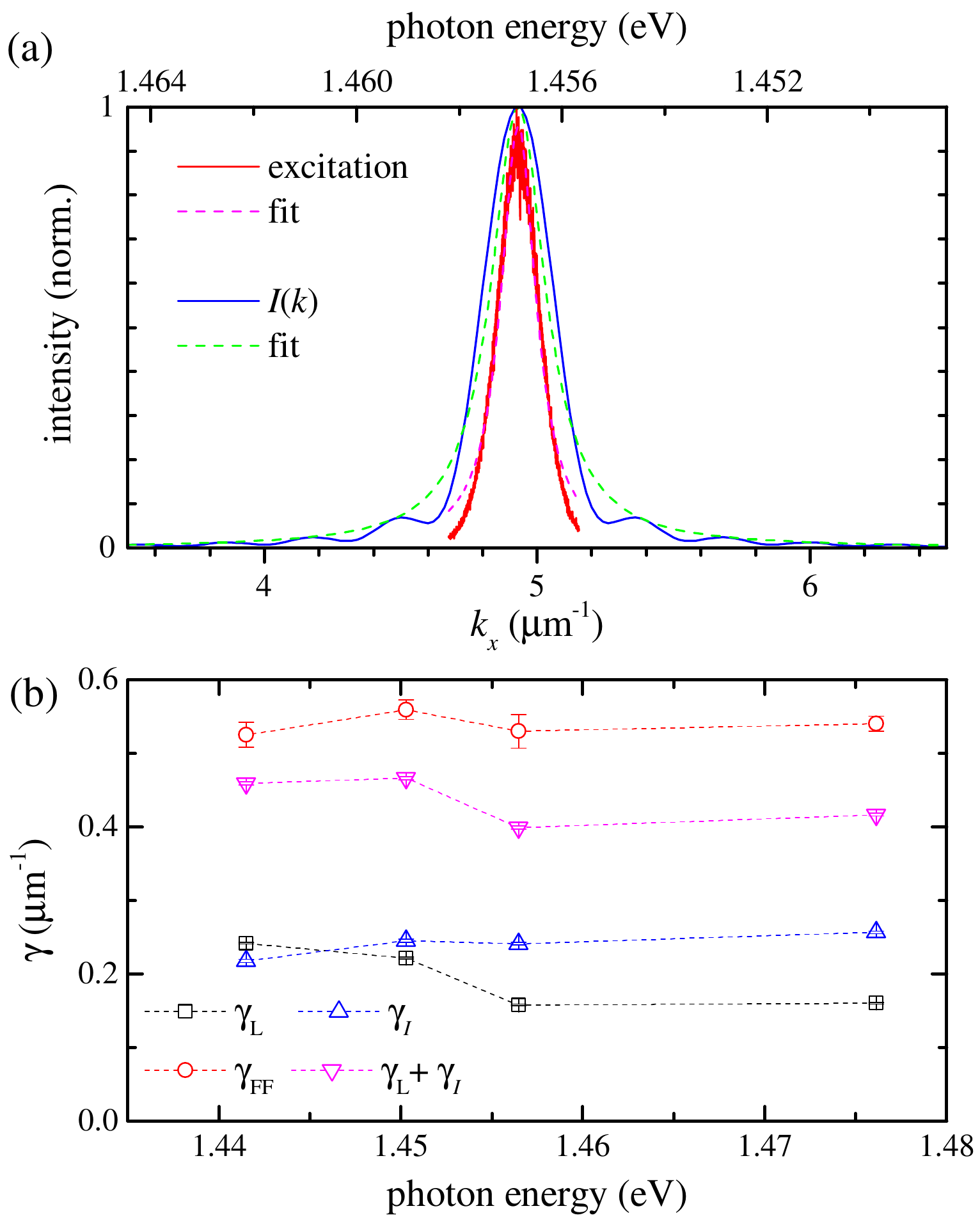}
	\caption{(a)\,Solid lines: excitation spectrum and $I(k)$; dashed lines: corresponding lorentzian fits. (b)\,FF width $\gamma_{\rm FF}$ in the $k_x$ direction interpreted as the sum of the excitation width $\gamma_{\text{L}}$ and the width $\gamma_{I}$ of $I(k)$. All widths are the FWHM of Lorentzian fits.} 
	\label{fig:FFResolution}
\end{figure}
In \Fig{fig:Dispersion}b in the main text, we show an example of the NF of the radiation losses along the waveguide. The data was obtained using the sum of the 100 subsequent frames of a video taken by a Sony DCR-TRV620E digital 8 camcorder in nightshot mode. The dark background was subtracted for all data shown. The NF emission has been derived as the difference between the emission profile along the waveguide and a background profile next to the waveguide. Each profile is averaged over 1.7\,\textmu m in the $x$-direction, as indicated in \Fig{fig:Dispersion}b with white and red dashed lines respectively. The $x$ axis is calibrated using the known distance between the couplers. We repeated the analysis for different excitation energies, and calculated the corresponding loss coefficients. The result is shown in the inset of \Fig{fig:Dispersion}a. The origin of these losses is discussed in \App{loss}. 

\subsection{\label{Sec:CarrierDiffusionLength} Carrier diffusion length}
The emission imaging allows to measure the carrier diffusion length on the photonic crystal membrane. In particular, we image the PL emission from the QDs for excitation about 3\,\textmu m sideways offset from WG, on the unstructured free-standing membrane, as shown in \Fig{fig:CarrierDiffusion}. The spatially resolved emission is extended compared to the excitation, due to carrier diffusion between excitation and recombination. In order to evaluate the diffusion length, we average over about 2.5\,\textmu m in the $y$-direction (see white dashed lines). The corresponding profile, given as solid white line, is fitted with a Gaussian profile, showing a standard deviation of about 1.4\,\textmu m, which represents the carrier diffusion length along the $x$ direction. Some of the carriers diffuse into the WG region, exciting QDs that couple to the guided mode, as is evident from the weak emission observed from the couplers.  

\subsection{\label{Sec:FFResolution}Far-field width} 
The far field profile (see \Fig{fig:Dispersion}c) has a finite FWHM, $\gamma_{\rm FF}$, in the $k_x$ direction. There are two contributions to the observed width: (i) the spectral width of the excitation laser, which translates via the PCWG dispersion into a FWHM $\gamma_{\rm L}$, and (ii) the spatial profile of the NF in the $x$ direction, which is given by an exponential decay of the field amplitude due to losses, delimited by the aperture size \Lp.  
The observed field amplitude is thus modelled, neglecting constant factors, as
\be E(x)=\left[\theta\left(x+\frac{\Lp}{2}\right)-\theta\left(x-\frac{\Lp}{2}\right)\right]\exp\left(-\frac{\alpha x}{2}\right),
\label{Eq:PinholeLossesFunc}\ee
with the Heaviside function $\theta$ and the aperture length \Lp. Fourier-transforming and taking the absolute square, we find the corresponding intensity in $k$-space is given by  
\be
I(k)=\frac{2}{k^2+\frac{\alpha^2}{4}}\left[\text{cosh}\left(\frac{\alpha}{2}\Lp\right)-\text{cos}\left(k\Lp\right)\right]\,. 
\label{Eq:FFResolution_kSpace}
\ee    
For large losses across the aperture, $\alpha \Lp\gg1$, the constant cosh term dominates and $I(k)$ is a Lorentzian, while for small losses, $\alpha \Lp\ll 1$, the cos term dominates, resulting in a sinc function. We fit \Eq{Eq:FFResolution_kSpace} with a Lorentzian to determine the equivalent FWHM $\gamma_{I}$. The laser spectrum, converted into $k_x$ using the linear dispersion of about $-27.3$\,\textmu m$^{-1}$ eV$^{-1}$ in the relevant range (see \Fig{fig:Dispersion}), and $I(k)$, together with the corresponding Lorentzian fits to determine $\gamma_{\rm L}$, and $\gamma_I$, respectively, are shown in \Fig{fig:FFResolution}a for \EL=1.4565\,eV, as used in \Fig{fig:Dispersion}b-c. The FF profile is then given by the convolution of (i) and (ii), which for Lorentzians keeps a Lorentzian, having a width given by the sum of the widths, here $\gamma_{\rm L}+\gamma_{I}$. The result of this analysis is shown in \Fig{fig:FFResolution}b for all measured \EL. Generally a good quantitative agreement is found. The remaining difference between $\gamma_{\rm L}+\gamma_{I}$ and $\gamma_{\rm FF}$ of about 0.1\,\textmu m$^{-1}$ could be related to a slight defocus of the FF imaging.

\begin{figure}
	\includegraphics[width=\columnwidth]{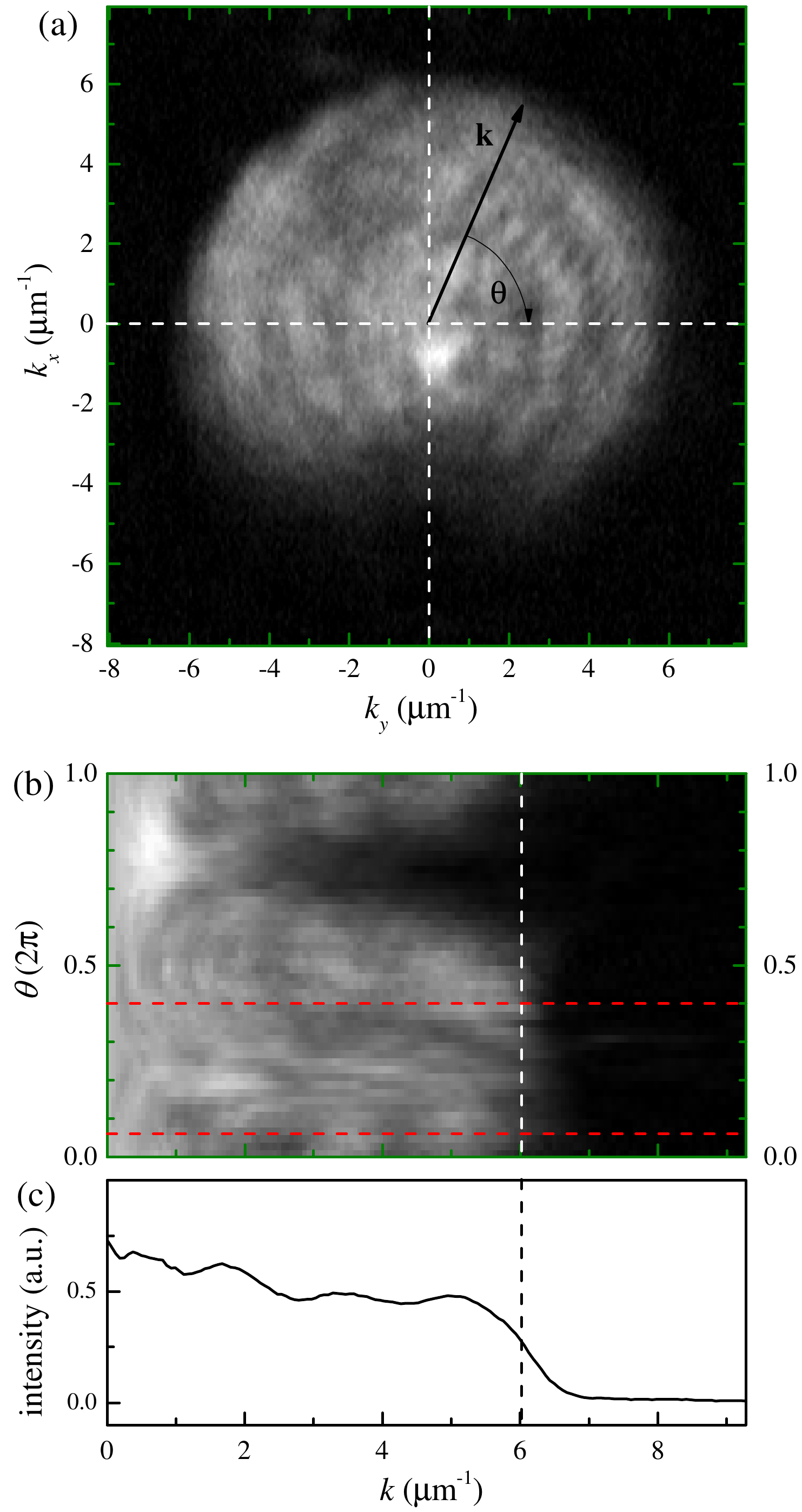}
	\caption{NA calibration of FF imaging. (a)\, Emission from top coupler for excitation into the bottom coupler at \EL=1.4102\,eV. (b)\, Polar representation of (a). (c)\,Radial profile of (b) averaged over of the $\theta$ range indicated by horizontal dashed lines in (b). Vertical dashed lines in (b) and (c) indicate the radius of the NA in $k$-space. Gray-scale as in \Fig{fig:CarrierDiffusion}.} 
	\label{fig:NACalibration}
\end{figure}
\subsection{\label{Sec:FFCalibration}Calibration of FF imaging}
In order to calibrate the in-plane wavevector in the FF imaging, we measured the FF of the emission of the top coupler when exciting into the bottom coupler at \EL=1.4102\,eV, as shown in \Fig{fig:NACalibration}a. To use the known NA of the objective for calibration, we determine the cut-off radius, using the polar coordinate representation of the data, as shown in \Fig{fig:NACalibration}b. We adjusted the center of the coordinate system to obtain a constant maximum radius, shown by the vertical dashed line. We average the $\theta$ range between the two horizontal dashed lines in \Fig{fig:NACalibration}b, obtaining the profile shown in \Fig{fig:NACalibration}c. We take the NA radius to be the value of ${k}$ at half step height, and calibrate this radius to $k_0 {\rm NA}=6.076$\,\textmu m$^{-1}$.
A relative error of a few \% rms of this calibration is estimated. 

\section{Simulation methods}
\label{app:simulation}
\subsection{Geometry and Parameters}
\label{mode_finding}

\begin{figure}
	\includegraphics[width=\columnwidth]{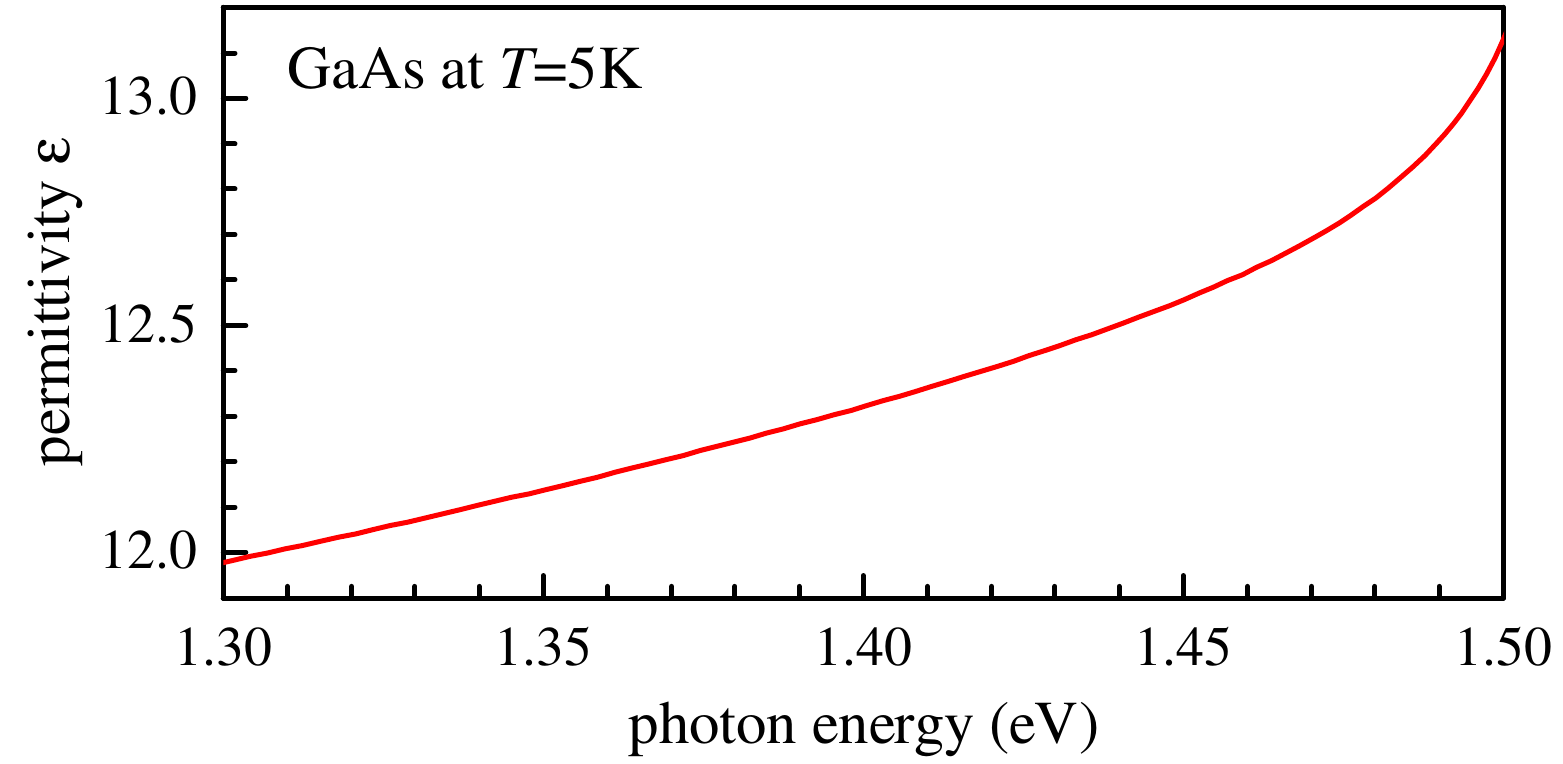}
	\caption{GaAs permittivity $\varepsilon$ as function of photon energy used in the calculations. Data from the material implementation of \Eq{equ:eps} in FDTD.}
	\label{fig:epsilon_gaas}
\end{figure}

FDTD calculations using the package Meep\,\cite{OskooiCPC10} were carried out to model the properties of the PCWGs under study. All calculations used a resolution of 24 points per lattice constant, and a cubic Yee-lattice. The permittivity  $\varepsilon$ of GaAs at low temperature $T=5$\,K including its dispersion \cite{GehrsitzJAP00} is given by
%
%
\begin{align} \label{equ:eps}
	\varepsilon(E) =& 5.965 + \\ &\frac{0.0304}{1.519^2 - E^2}+\frac{33.1494}{2.692^2 - E^2}+ \frac{0.00238}{0.0334^2 - E^2}\nonumber\,,
\end{align} 
with the photon energy $E$ in units of eV. This permittivity was implemented in the FDTD dispersion model as shown in \Fig{fig:epsilon_gaas} over the relevant energy range.

\begin{figure}
	\includegraphics[width=\columnwidth]{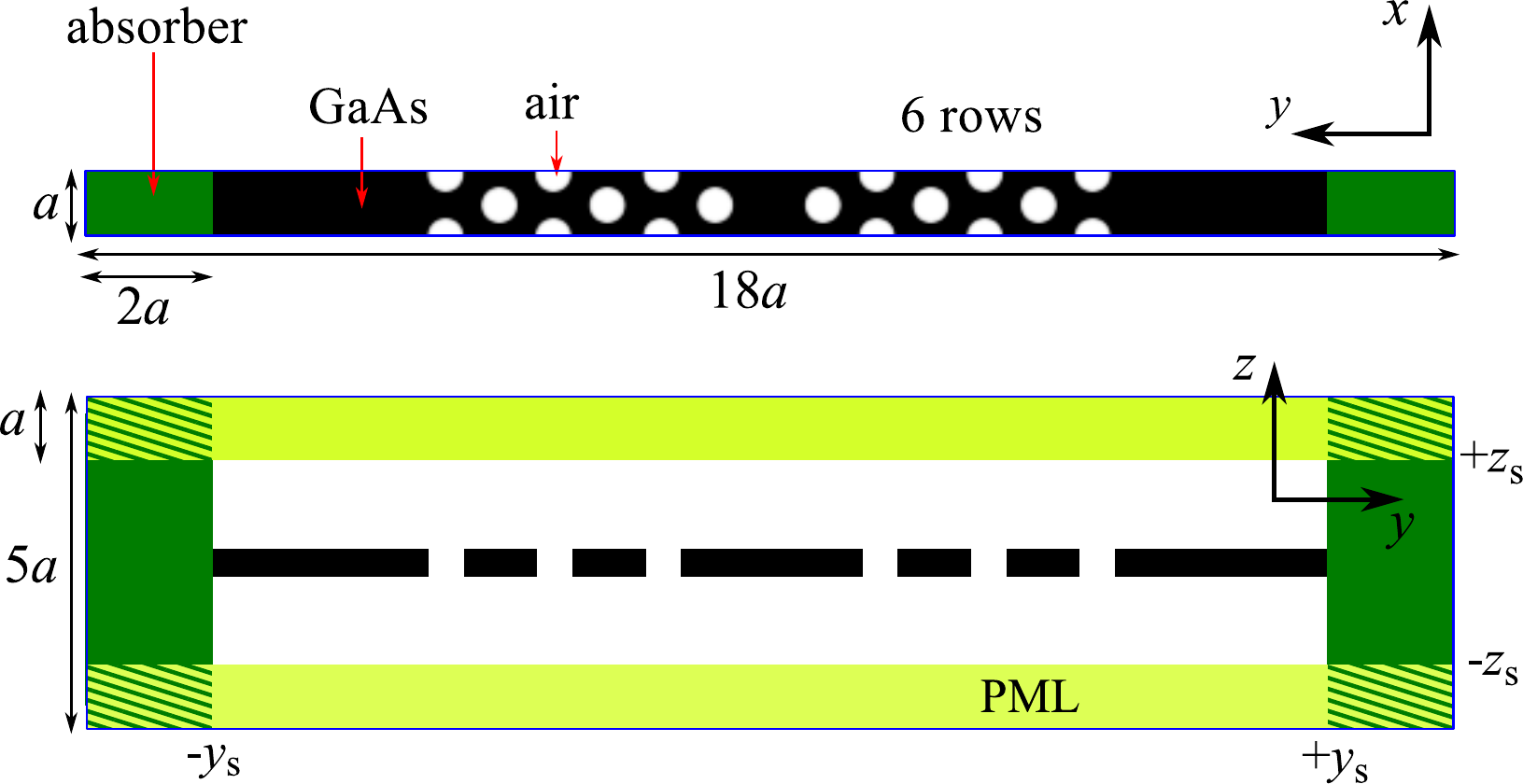}
	\caption{Sketch of the simulation volume used in the band structure calculations, covering one unit cell of the PCWG in $x$-direction.}
	\label{fig:single_cell_diagram}
\end{figure}

Single unit cell simulations were carried out with periodic boundary conditions along $x$ and a current source at $z=0$ and a given position in $x,y$. Initially these calculations made use of the nominal sample parameters: lattice constant $a = 260$\,nm, hole radius $r = 0.24a$ and slab height $h = 0.4808a=125$\,nm. Perfectly matched layers (PMLs) were placed on the out-of-plane simulation facets at $\pm \zs$ and adiabatic absorbers \cite{OskooiOE08} on the in-plane facets at $\pm \ys$ to absorb light tunneling through the cladding. Six layers of air holes were placed either side of the waveguide, matching the sample. A sketch of the geometry covering one unit cell of the PCWG is shown in \Fig{fig:single_cell_diagram}. 

To calculate the WG mode dispersion, we implemented periodic boundary conditions with a phase factor over a single PCWG unit cell, and extracted the mode resonances from the time-domain Fourier transforms as function of the phase. 

\begin{figure}[b]
	\includegraphics[width=\columnwidth]{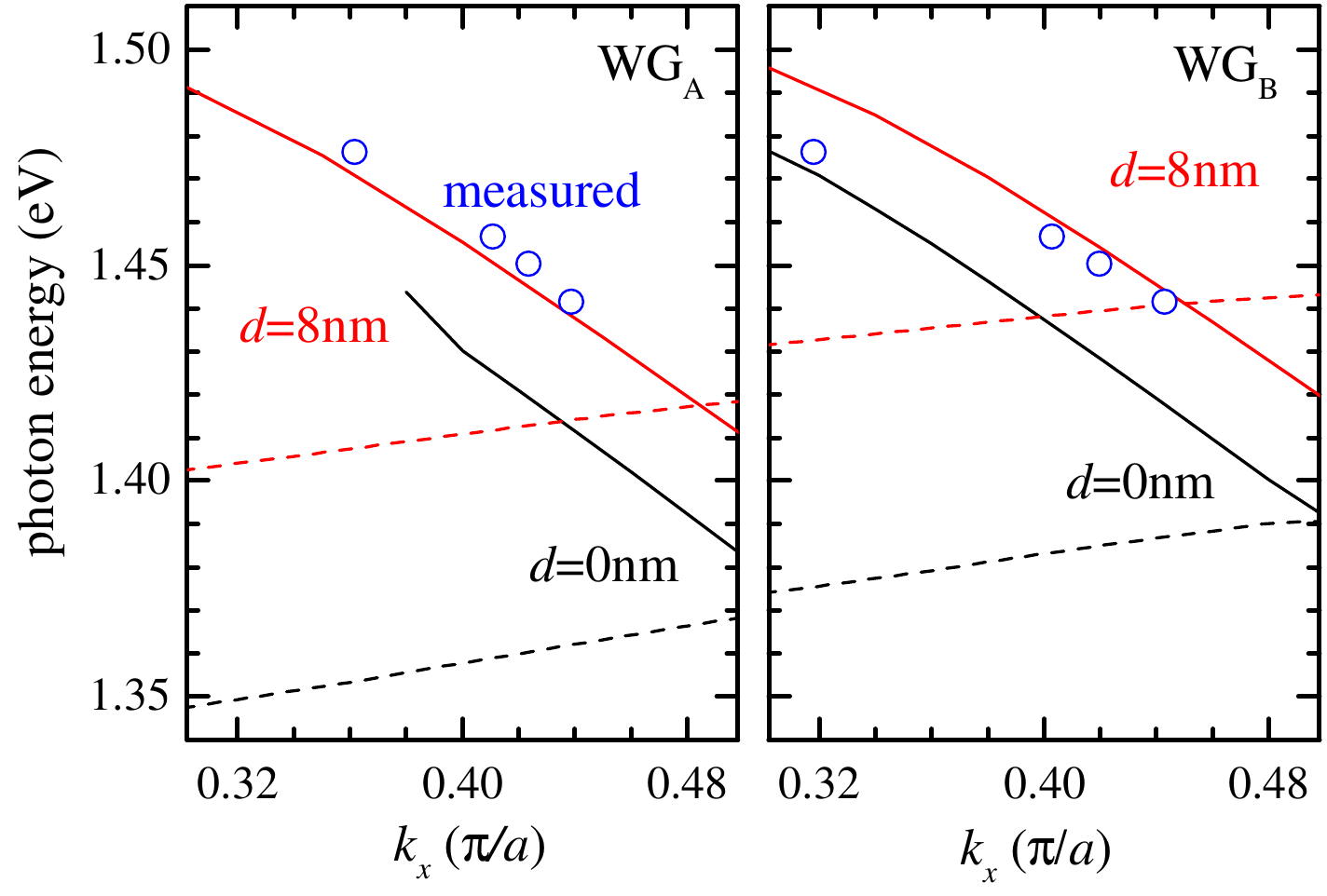}
	\caption{Matching simulations to the measured mode dispersion for \WGA\ (left) and \WGB\ (right). Solid lines: fundamental mode; dashed lines: higher-order mode; black lines: $d=0$\,nm; red lines: $d=8$\,nm; Circles: measured (see \Fig{fig:Dispersion}).}
	\label{fig:etch}
\end{figure}

Simulations with the nominal structural parameters showed significant deviations from the measured dispersion. As discussed in the main text, a single parameter $d$ was introduced, describing the thickness of material removed from all surfaces of the structure, for example by etching during fabrication or subsequent oxidation. Simulations were carried out with the radius of the holes expanded by $d$, and the height of the slab reduced by $2d$. We found that $d$ around 7-8\,nm produces a good match with experiment, see \Fig{fig:etch}, and used $d=8$\,nm for the remaining calculations, for both investigated WGs.

\subsection{PCWG loss}
\label{loss}

A significant number of QDs measured are in resonance with WG modes with propagation wavevectors $k_x$ within the light cone, $k_x < 2\pi \nu c/a$ with the mode eigenfrequency $\nu$  and the speed of light $c$. Modes in this part of the dispersion emit into free space and thus experience propagation loss, even in an ideal structure. In addition to this out-of-plane loss, all modes will experience an in-plane loss due to the finite extension of the lateral PC, allowing light to tunnel through the PC cladding either side of the WG.

\begin{figure}
	\includegraphics[width=\columnwidth]{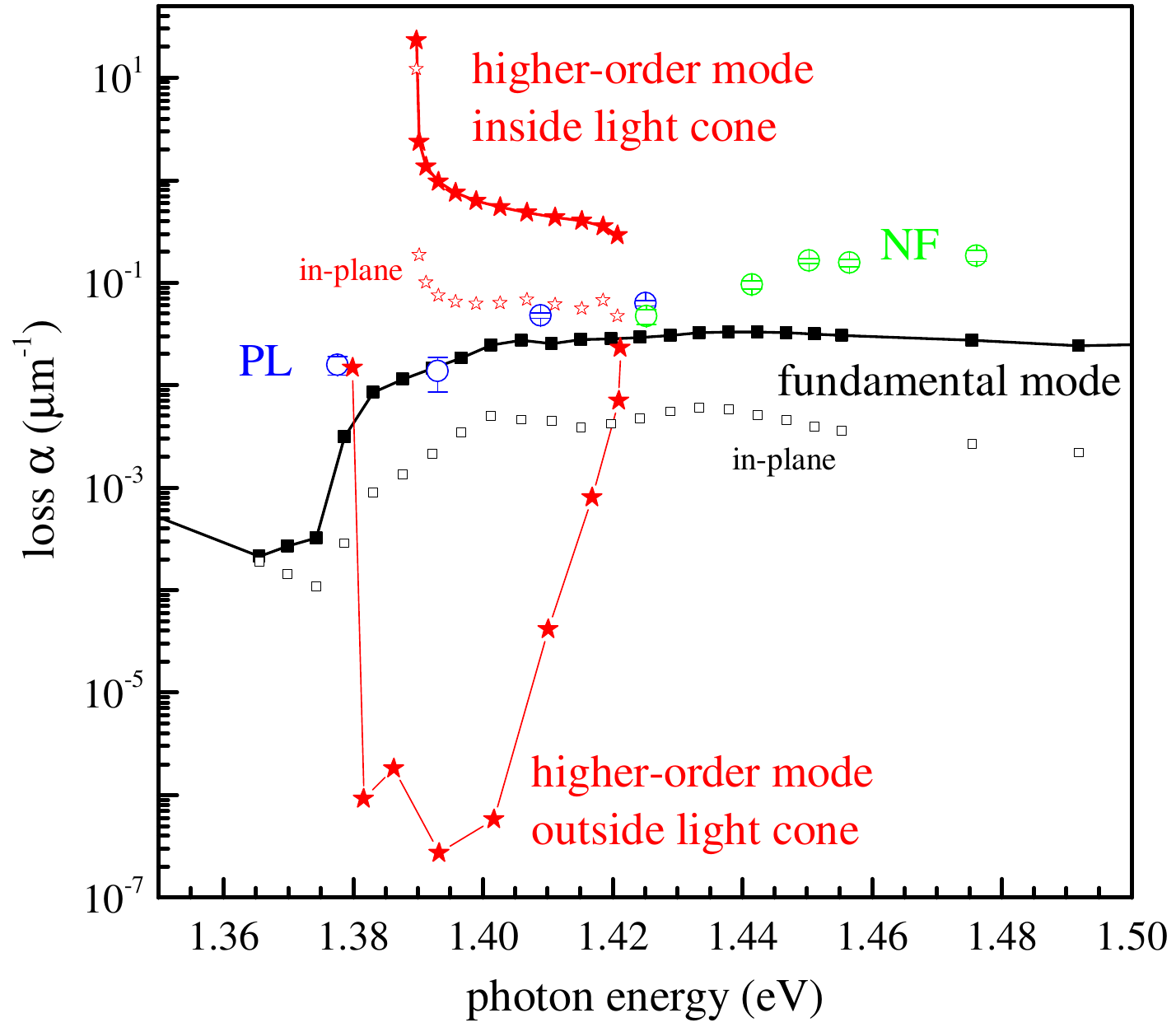}
	\caption{Simulated propagation loss of the fundamental mode (filled black squares) and higher order mode (filled red stars). The in-plane losses are given as empty symbols. Blue circles: loss measured from PL. Green circles: loss measured from NF.}
	\label{fig:loss}
\end{figure}

\begin{figure*}
	\centering
	\includegraphics[width=\textwidth]{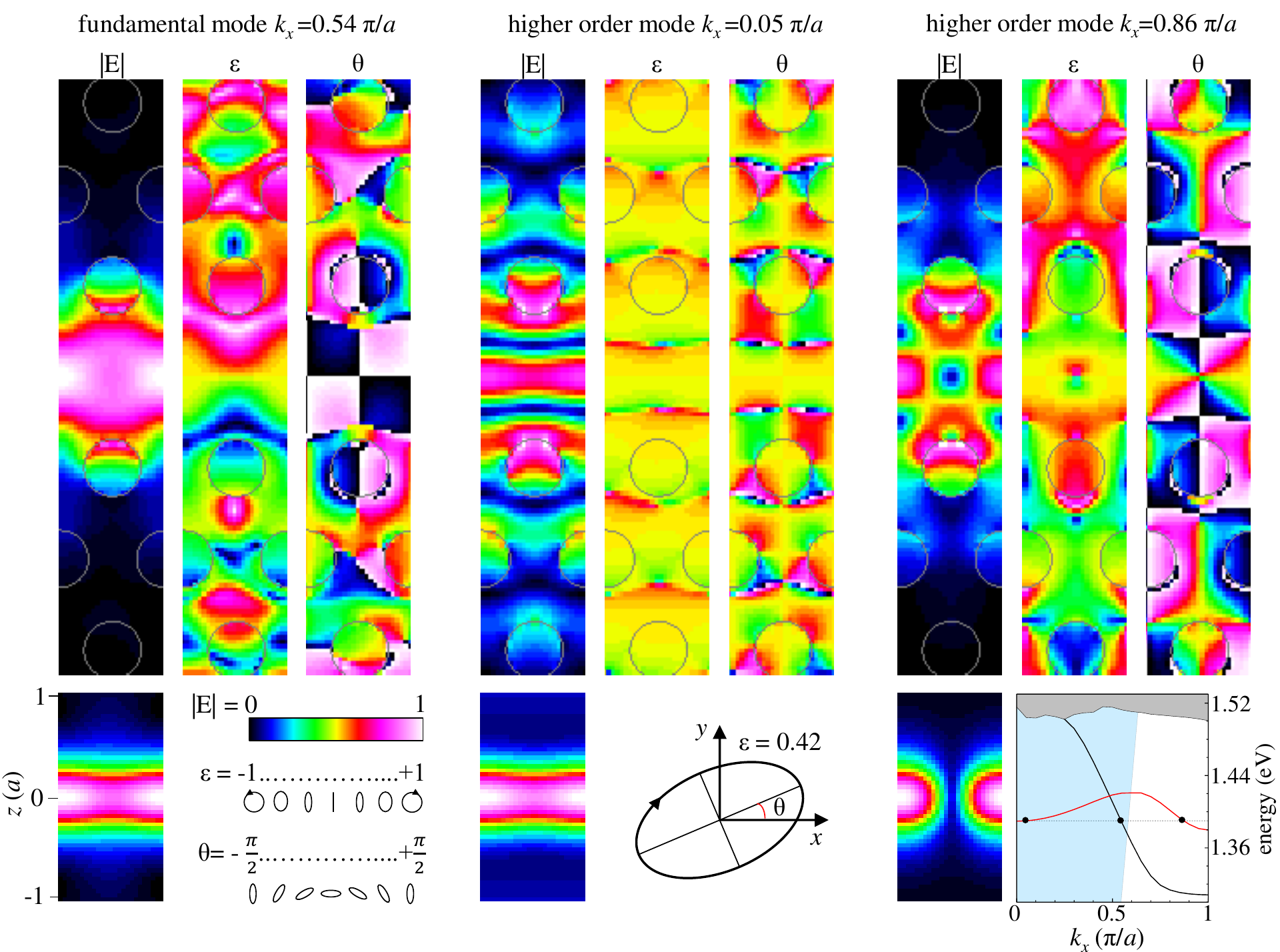}
	\caption{Absolute value of the electric field amplitude $|E|$, polarisation ellipticity $\varepsilon$, and orientation $\theta$, in the plane $z=0$, for $\WGA$. The hole structure is superimposed for reference. $\theta$ is defined according to the sketch shown in the central panel. The source energy is $E=1.39$\,eV, which corresponds to the indicated propagation wavevectors, according to the inset in the right panel, which are imposed by the phase of the boundary condition. $xz$ sections of $|E|$ at $y=0$ are shown below the corresponding $xy$ sections.}
	\label{fig:FieldProfile}
\end{figure*} 

The loss was simulated with the method described in \App{mode_finding}, choosing the symmetry of the simulation and sources either even or odd to excite only one of the two WG modes (see \App{Sec:Grating Couplers}). The simulation geometry is as sketched in \Fig{fig:single_cell_diagram}, with the height extended to $10a$ and the PML thickness to $1.5a$, in order to allow for better separation of the out-of-plane loss from the in-plane propagation. We use a source with a Gaussian time dependence of 29\,fs standard deviation, truncated at 5 standard deviations either side of the peak. To maximize excitation of the WG mode of interest, the source center frequency was set to the WG mode frequency at the simulated $k_x$, as calculated in \App{mode_finding}. The simulation was run for 1000 time-steps, ending about $870$\,fs after the source is switched off, at which point the remaining field of non-guided modes is negligible, as is evidenced by a stable field distribution, apart from a global oscillating and decaying pre-factor $\exp(-i\omega t)$ with the complex mode frequency $\omega(k_x)$. The Poynting vector of this field was then evaluated, providing the power flux density of the selected WG mode. This distribution was used to calculate the flux $\Fx$ through the $+\xs$ plane, representing the propagating flux along the PCWG, and the flux $\Fz$ through the $\pm \zs$ planes representing the out-of-plane loss, and $\Fy$ through the $\pm \ys$ planes, representing the in-plane loss. The transmission coefficient per unit cell is then given by $T = \Fx/(\Fz + \Fy + \Fx)$, from which the loss coefficient $\alpha=-\ln(T)/a$ is determined, as shown in  \Fig{fig:loss} for the two WG modes as a function of energy. The loss coefficient considering in-plane loss only (using $\Fz=0$) is also given. The even (higher order) mode has a dispersion showing a maximum, and thus presents two $k$-vectors for a given energy over a significant range. One branch corresponds to the part of the even mode inside the light cone, showing high loss, dominated by out-of-plane loss, while the other is outside the light cone and shows only the small remaining in-plane loss due to the tunnelling through the cladding. The odd (fundamental) mode shows increasing loss with increasing energy. Below 1.38\,eV, the mode is outside the light cone, so that the loss is strongly reduced to the in-plane loss only. The measured loss is indicated by the circular data points, and show a good agreement with the calculated loss of the odd mode. The additional loss in the measurements above 1.43\,eV is attributed to absorption in the wetting layer, not taken into account in the simulations.

\begin{figure*}
	\includegraphics[width=\textwidth]{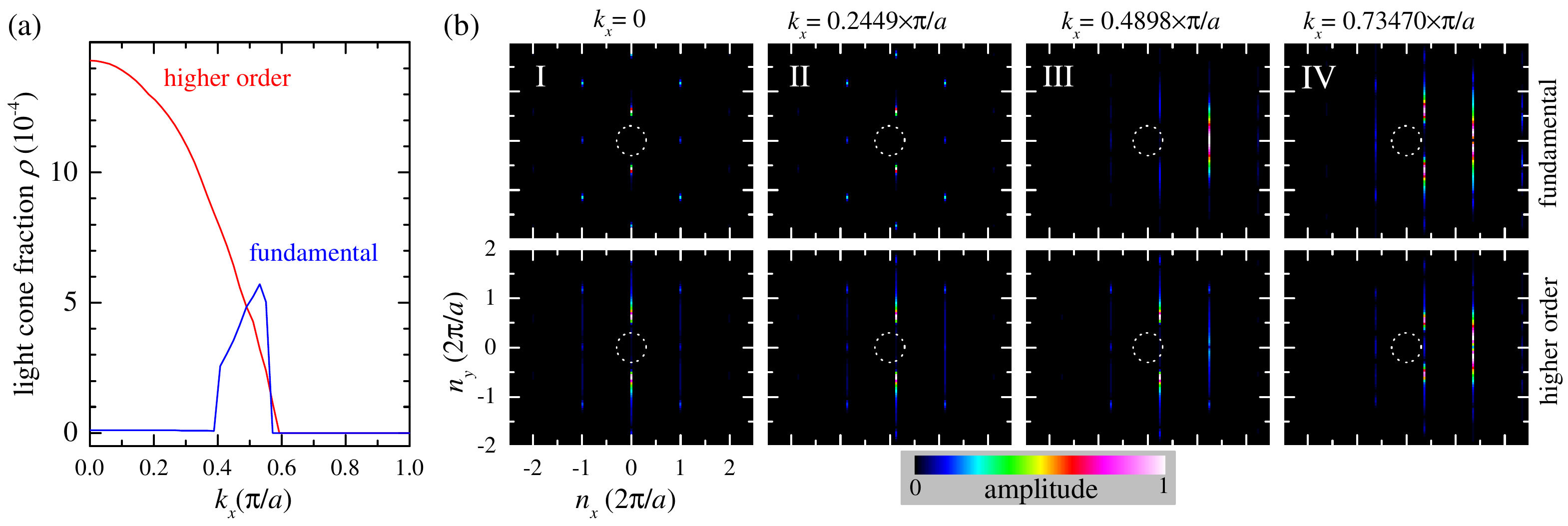}
	\caption{Fourier analysis of WG mode fields. (a) Fraction $\rho$ of intensity $|\tilde{H}_z(\bn)|^2$ inside the light cone \LC, for the fundamental (blue line) and the higher order (red line) mode, versus Bloch wavevector $k_x$. (b) $|\tilde{H}_z(\bn)|$ of the fundamental (upper panels) and higher order (lower panels) modes at specific $k_x$ as indicated. The white circles represent the light cone \LC.}
	\label{fig:sft}
\end{figure*}

From the same simulations we also obtain the mode field amplitude and polarisation. In \Fig{fig:FieldProfile}, the absolute value of the electric field amplitude $|E|$, the ellipticity $\varepsilon$ and the orientation $\theta$ of the polarisation ellipse are shown in the plane $z=0$, for $\WGA$. Note that for $z=0$ the $z$ field component is zero. Orientation and ellipticity are defined as 
\be
\begin{aligned}
	&\theta=\frac{1}{2}{\rm arctan}\left(\frac{S_2}{S_1}\right)\\
	&|\varepsilon|=\sqrt{\frac{2}{\sqrt{S_1^2+S_2^2}+1}-1}\\
	&\varepsilon=|\varepsilon|{\rm sign}(S_3)\hspace{0.2cm},
\end{aligned}
\label{Eq:EllipticityOrienatiton}
\ee  
with the Stokes parameters $S_1=(|E_x|^2 - |E_y|^2)/S_0$, $S_2=2 \text{Re}(E_x^* E_y) / S_0$ and $S_3=2 \text{Im}(E_x^* E_y) / S_0$ and $S_0=|E_x|^2 + |E_y|^2$. 
We can clearly see the spatial extension of the modes in their amplitude, and their circular and linear points in their ellipticity. The fundamental mode is linear in the center (yellow in $\varepsilon$), and circular of opposite helicity (black and white) in the center of the unit cell along $x$, close to the first holes. Other linear and circular points are present, but at much reduced mode field amplitudes. The higher order mode at a small $k_x=0.05\,\pi/a$ is dominated by linear polarization, and at large $k_x=0.86\,\pi/a$ has circular points at the edge of the unit cell, at lower field strengths. This is a consequence of the temporal symmetry that ensures that no component of circular polarisation can exist at the band edge points $k_x = 0$ and $k_x = \pm \pi/a$ in this type of waveguide \cite{LangPTRSA16}. Thus in the proximity of these $k_x$ one finds either a reduced component of circular polarisation and/or a reduction in the electric field strength in the regions with circular polarisation.

To give some insight into the radiative loss of the modes, 2D spatial Fourier transforms were carried out on the fields of the guided modes. The fields in the $z$=0 plane of the slab were used. In this plane, symmetry ensures that the three non-zero field components are $E_x(\bm r)$, $E_y(\bm r)$ and $H_z(\bm r)$. The profile of $|H_z(\bm r)|^2$ appears nearly identical to $|E_x(\bm r)|^2 + |E_y(\bm r)|^2$, and we choose to use for the following discussion $\Hzt(\bn)$, the Fourier transform with wavevector $\bn$. Components at $|\bn|<k_0$, inside the light cone \LC, can couple to free space modes, having an intensity fraction  $\rho=(\int_\LC \xi(\bn) {\rm d}\bn)/(\int \xi(\bn) {\rm d}\bn)$, with $\xi(\bn)=|\Hzt(\bn)|^2 + |E_x(\bn)|^2 + |E_y(\bn)|^2$, which is indicative of the loss rate per unit time\,\cite{AkahaneOE05}. For this calculation, the mode fields were extracted from an eigensolver \cite{JohnsonOE01}, using $\varepsilon = 12.25$, and $d=0$\,nm. The fraction $\rho$ is given in \Fig{fig:sft}a for both WG modes as function of the Bloch wavevector $k_x$, and the distribution of $|\Hzt(\bn)|$ is shown in \Fig{fig:sft}b for selected $k_x$. 

\begin{figure}
	\includegraphics[width=\columnwidth]{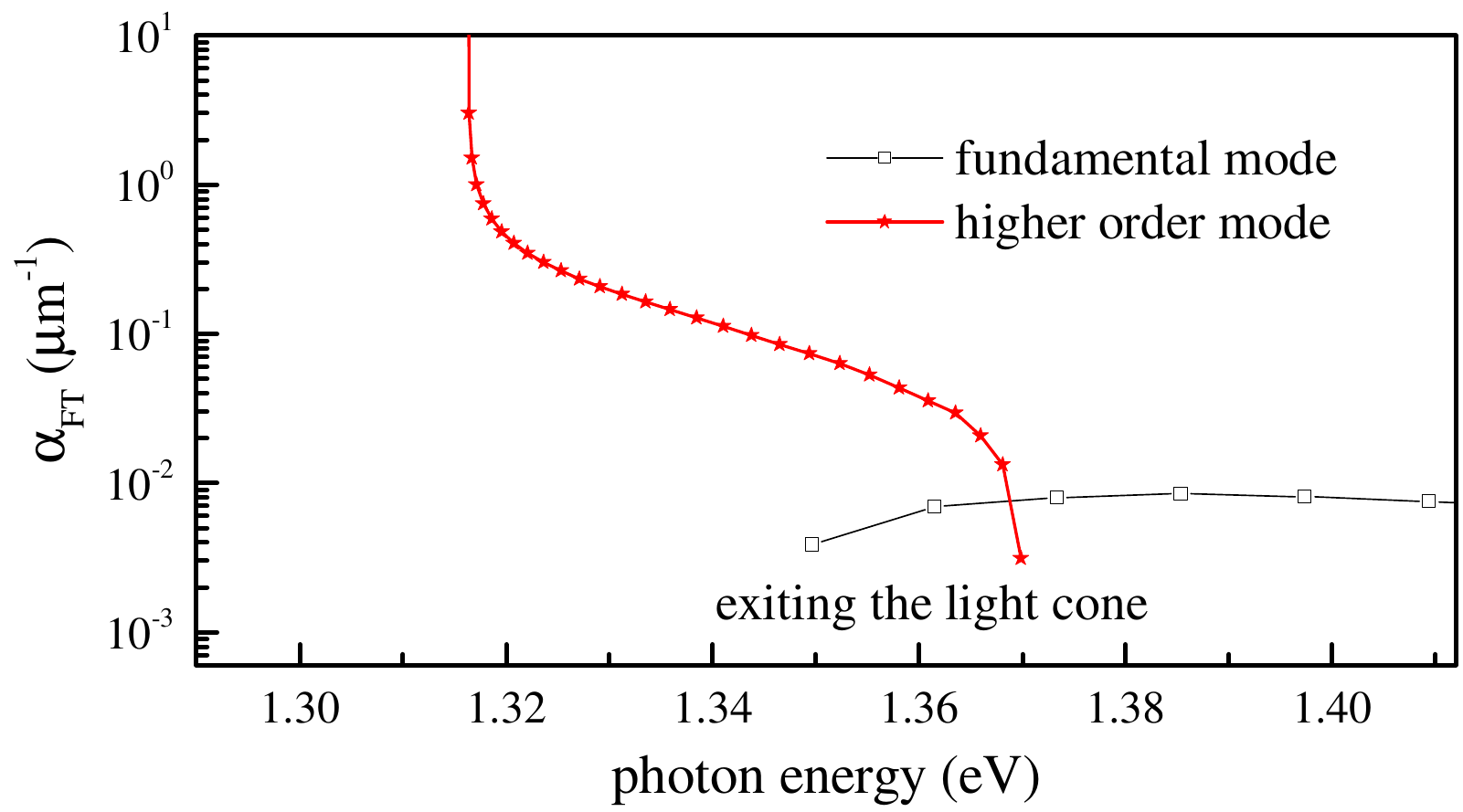}
	\caption{Estimated loss \Aft\ of the two WG modes using \Eq{Eq:fourier_loss} based on the fraction $\rho$ inside the light cone, shown in \Fig{fig:sft}.}
	\label{fig:fourier_loss}
\end{figure}

The fundamental mode for small $k_x$ (see \Fig{fig:sft}b panels I and II) is in the continuum above the bandgap of the PC cladding, and thus able to propagate in the $\pm y$ directions, leading to a small extension in $n_y$. This concentration reduces the overlap with the light cone, hence the low $\rho$ in \Fig{fig:sft}a in this region. For higher $k_x$ (see panel III) the mode enters the bandgap, confining it in $y$ and thus broadening it in $n_y$. This results in an increased $\rho$, until $k_x$ leaves the light cone (panel IV). In contrast, the higher order mode (see bottom row of \Fig{fig:sft}b) remains inside the bandgap for all $k_x$, so that $\rho$ is decreasing monotonically with increasing $k_x$, reaching zero at the edge of the light cone, $k_x=k_0$.

In order to estimate the propagation loss coefficient  $\Aft$ per unit distance, the fraction $\rho$ is scaled as
\be
\Aft=\frac{1}{v_{\rm g}}\frac{c k_z}{h k_0}\rho\,.
\label{Eq:fourier_loss}
\ee
In this expression, the factor $(c k_z)/(h k_0)$ is the attempt rate with which the light is emitted, with the emission fraction $\rho$ per attempt. The factor 1/$v_{\rm g}$, with the group velocity $v_{\rm g}$, converts the resulting loss per unit time into the loss per unit propagation length.  
The resulting loss is given in \Fig{fig:fourier_loss}. This approach only accounts for out-of-plane losses and assumes an infinite WG without absorption and disorder. The loss thus becomes exactly zero outside the light cone, beyond the lower end of the curves in \Fig{fig:fourier_loss}. Notably, this estimate of loss is qualitatively reproducing the one calculated via FDTD (see \Fig{fig:loss}) for both modes, and is quantitatively about a factor of 3 lower.

\subsection{Coupler efficiency}\label{Sec:Grating Couplers}

\begin{figure*}
	\includegraphics[width=\textwidth]{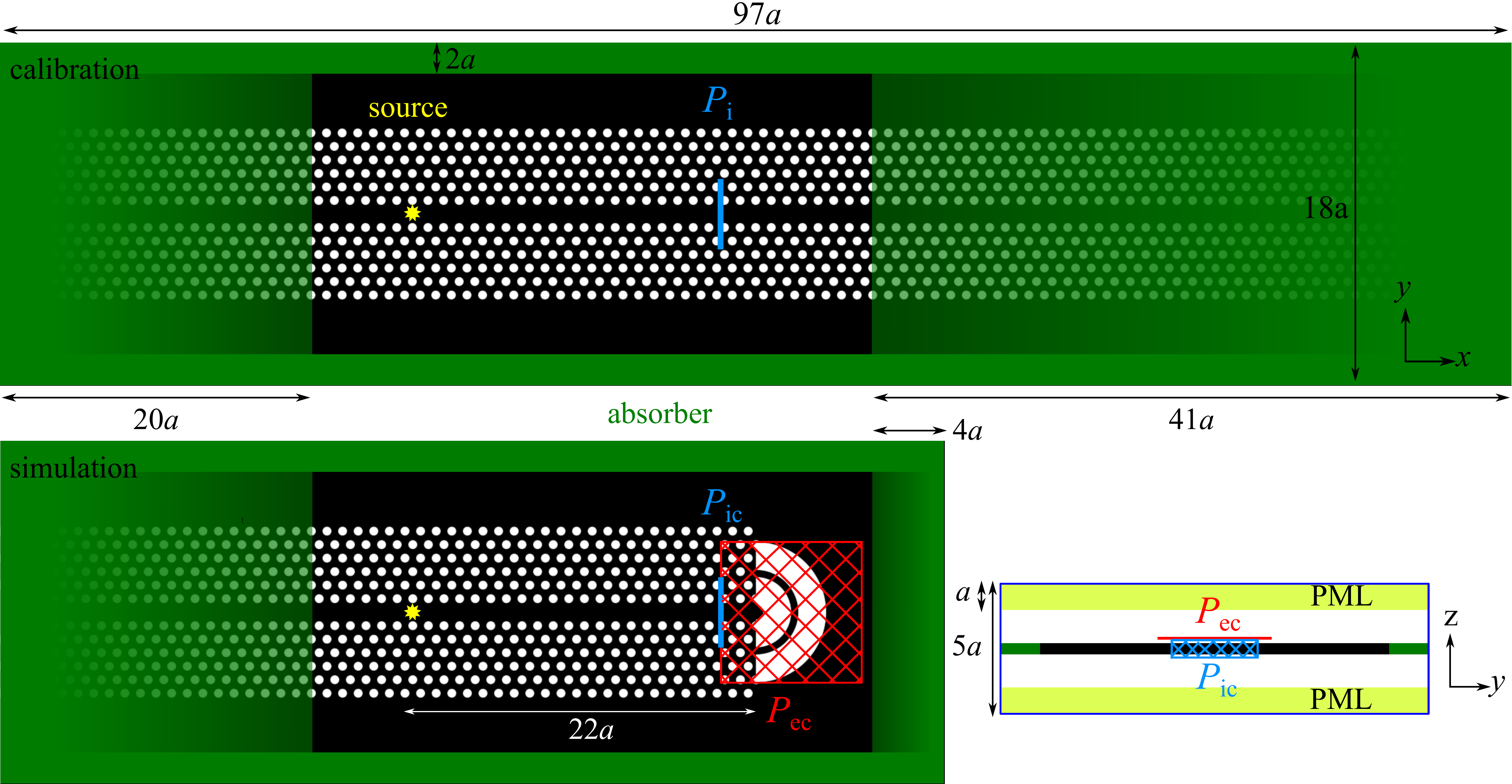}
	\caption{Sketch of the FDTD simulation volumes used to determine coupler efficiency and reflectivity, consisting of a calibration simulation (top), and a coupler simulation (bottom). The green areas indicate absorbers. The blue line indicates the plane that measures the powers \Pin\ and \Pic. The red hatched region indicates the area over which the power emitted by the coupler in measured. On the bottom right a view along the WG is given, stretched by 20\% along $z$ for clarity. For the even-mode calculations, the simulation volume is expanded by $5a$ on each $x$ facet. The radial bars supporting the grating were not included in the simulation.}
	\label{fig:grating_sim_map}
\end{figure*}

FDTD simulations were used to determine the efficiency of the couplers. A sketch of the simulation volume used is shown in \Fig{fig:grating_sim_map}. In order to calculate the efficiency and reflectivity of the coupler, two simulations are used. The coupler simulation contains the actual structure, and a calibration simulation is identical except that the coupler is removed and replaced by a continuation of the PCWG. The power in the PCWG mode travelling towards the coupler, called the input power \Pin, is determined by the calibration simulation, given by the power through the flux plane indicated in blue in \Fig{fig:grating_sim_map}. After determining the power \Pic\ through the same flux plane in the coupler simulation, the coupler reflectivity is calculated as $\Rc=1-\Pic/\Pin$. Using the power radiated out of the coupler, \Pec, through the flux plane indicated in red in \Fig{fig:grating_sim_map}, located $a/20 = $ 13\,nm above the slab, the coupler efficiency is calculated as $\etac=\Pec/\Pin$. Note that due to the inversion symmetry in $z$, the same power is emitted towards both sides of the slab WG. 

In the simulations, an electrical dipole source (yellow star in \Fig{fig:grating_sim_map}) with a time-varying current given by $[j_x(t), j_y(t)] \propto \mathbf{d} \exp(-i\omega t-t^2/(2\Delta_t^2))$ is used, with the time $t$, the central frequency $\omega$, the duration $\Delta_t$, and the dipole vector $\mathbf{d}$. Note that MEEP implements such a source using discrete time derivatives of Gaussians for better performance. The source frequency was set to $\omega = 1.8395\,c/a$, at the centre of the measured QD distribution ($\hbar\omega \approx 1.3936$\,eV). The source has a duration of $\Delta_t=50\,a/c\approx 43$\,fs standard deviation, which results in a frequency standard deviation of $\Delta_\omega=2\pi/\Delta_t \approx 10$\,meV$/\hbar$, covering the range of QD energies measured experimentally. To convert the simulated fields from time-domain to frequency-domain, we use MEEPs flux plane function\,\cite{OskooiCPC10}.

The source and simulation symmetry were chosen to select for either even or odd modes. For the odd mode, the source was polarised along the $y$-direction ($\mathbf{d} = [0, 1]$), and placed at coordinates $[0.057, 0]a$ relative to the centre of the unit cell (see \Fig{fig:FieldProfile}). For the even mode, two in-phase $x$-polarised sources are used ($\mathbf{d} = [1, 0]$) placed at $[0.057, \pm 0.31]a$. The symmetries were exploited to gain a factor of 4 reduction in simulation time and memory, using an even mirror plane at $z=0$, and a mirror plane at $y=0$ with a parity matching the simulated mode \cite{OskooiCPC10}.

The simulation size is $97a$ in $x$, $18a$ in $y$, and $5a$ in $z$ direction, as shown in \Fig{fig:grating_sim_map}. Thick adiabatic absorbers covering the reflecting simulation boundaries in $x$ direction were found to be necessary,  since PCWG modes, specifically at low group velocity, are easily reflected from the spatially varying absorption in the absorbers.  The absorbers are implemented via an electric and magnetic conductivity, and we used a scaling of this conductivity proportional to the sixth power of the depth into the absorber, which provided lower reflections compared to using second and forth power. For the odd mode, in the calibration simulation, the absorber thickness in the positive $x$-direction was set to $41a$. The negative $x$ direction simulation boundary is less critical, as any reflections from this boundary will match between the calibration and coupler simulations, so that only the weaker reflection of the coupler reflection in the coupler simulation needs to be sufficiently suppressed. We used a thickness of $20a$. The other simulation facets do not have photonic structures intersecting them, and thus can be treated with shorter absorbers. We used $2a$ in the $\pm y$ directions, and PMLs of thickness $a$ in $z$. In the coupler simulation, there is no photonic lattice at the $+x$ boundary, and an absorber of $4a$ thickness was used. The frequency window of interest includes slow group velocity regions of the even mode, making them even more sensitive to being reflected by absorbers. In the even-mode simulations, the absorber thickness on the $\pm x$ boundaries were increased by $5a$. The absorber strength was set to 1/2 (1/18) of MEEP's default \cite{OskooiCPC10}, for the odd (even) mode, respectively. We found that \Pin\ varied by about 0.7\% in calibration simulations changing the absorber thickness on the $\pm x$ facets to $15a$. Furthermore, displacing the source by 1 unit cell along the waveguide, altered the fluxes by about 0.7\%. All relevant calculations were repeated with $5a$ less absorber depth on the $\pm x$ facets to ensure the accuracy of the results, and a change the measured fluxes of a few \% was observed.

\begin{figure}
	\centering
	\includegraphics[width=\columnwidth]{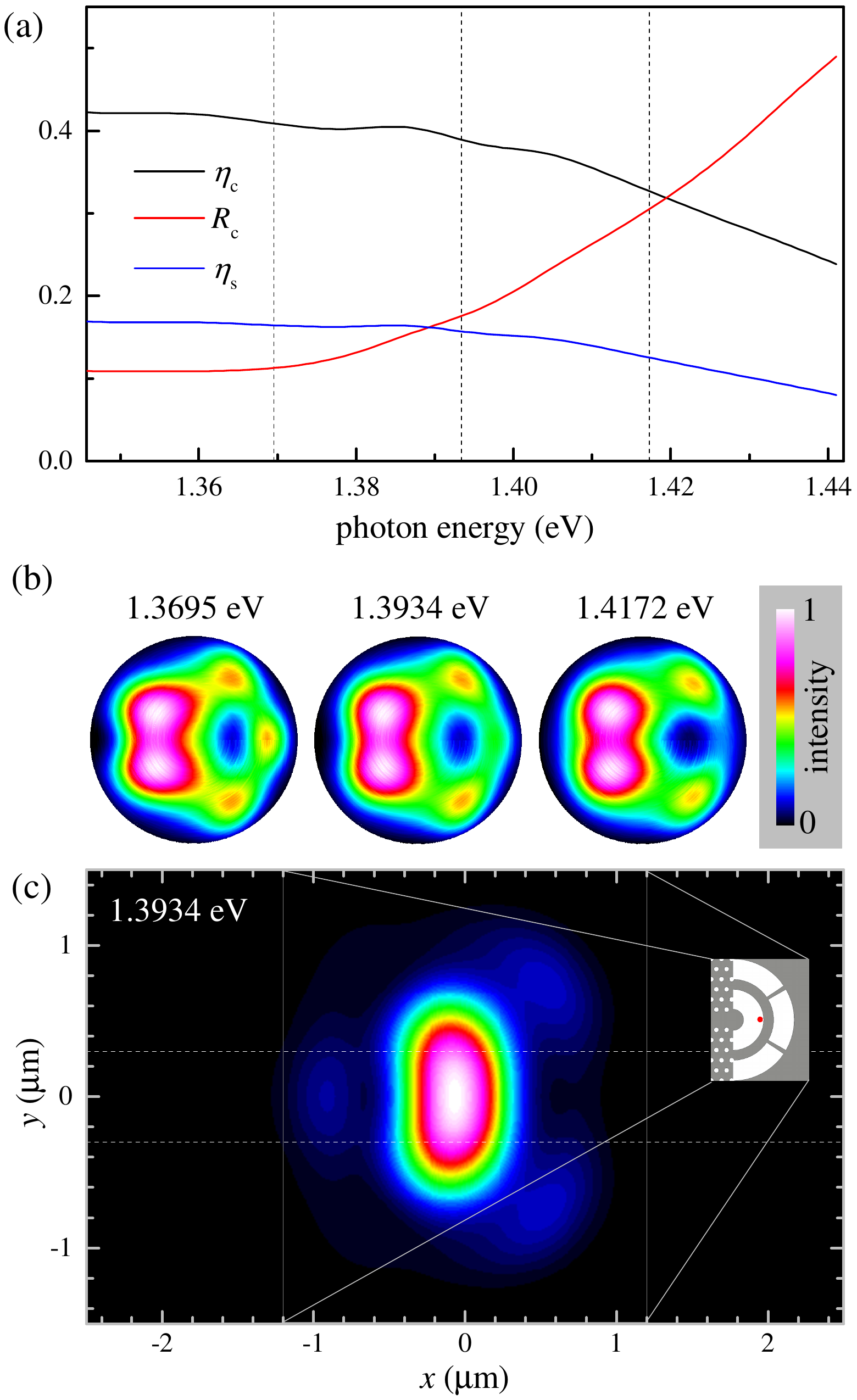}
	\caption{(a)\,Simulated efficiency \etac\ (black line) and reflectivity \Rc\ (red line) of the grating coupler for the odd (fundamental) WG mode as a function of energy. The blue line indicates the coupler efficiency into the spectrometer \etas, taking into account the microscope NA and the spectrometer slit. (b)\,Normalized far-field intensity of the coupler for three selected energies. The radius of the plots is $k_0$. (c)\,Simulated near-field image at 1.3934\,eV for an objective of 0.85NA, imaged with a magnification of 31.3 onto the spectrometer slit. The scale given refers to the size at the sample. The size of the spectrometer input slit is shown by the dashed lines. Inset: sketch of the coupler; the red dot indicates $x=y=0$\,\textmu m.}
	\label{fig:grating}
\end{figure}

The simulated efficiency \etac\ and reflectivity \Rc\ is shown in \Fig{fig:grating}a as a function of energy for the fundamental WG mode. Note that the coupler efficiency of about 40\% is referring to a single-sided emission, and that double sided efficiencies are twice as high considering the reflection symmetry in $z$.

We note that we find a reflectivity \Rc\ of 10-20\% for the odd mode in the relevant QD energy range (see \Fig{fig:SpectralImages}), while the measured circularity provides an upper limit (see \App{sec:GratingReflection}) of about 5\%. There are two main aspects contributing to this difference:
(i) the singly reflected light propagates up to $2L$ further. Using the typical propagation losses of $0.02/$\textmu m (see \Fig{fig:LossesPlots}) over a distance of $L$ we find a transmission factor of 0.6. (ii) \Pic\ is measured close to the coupler, and thus \Rc\ contains contributions from non-guided modes. For the even mode, the light reflected is furthermore distributed between the two branches inside and outside of the light cone, with the former having strong propagation losses.
Both aspects reduce the reflectivity relevant for the measured circularity.

\begin{figure}
	\includegraphics[width=\columnwidth]{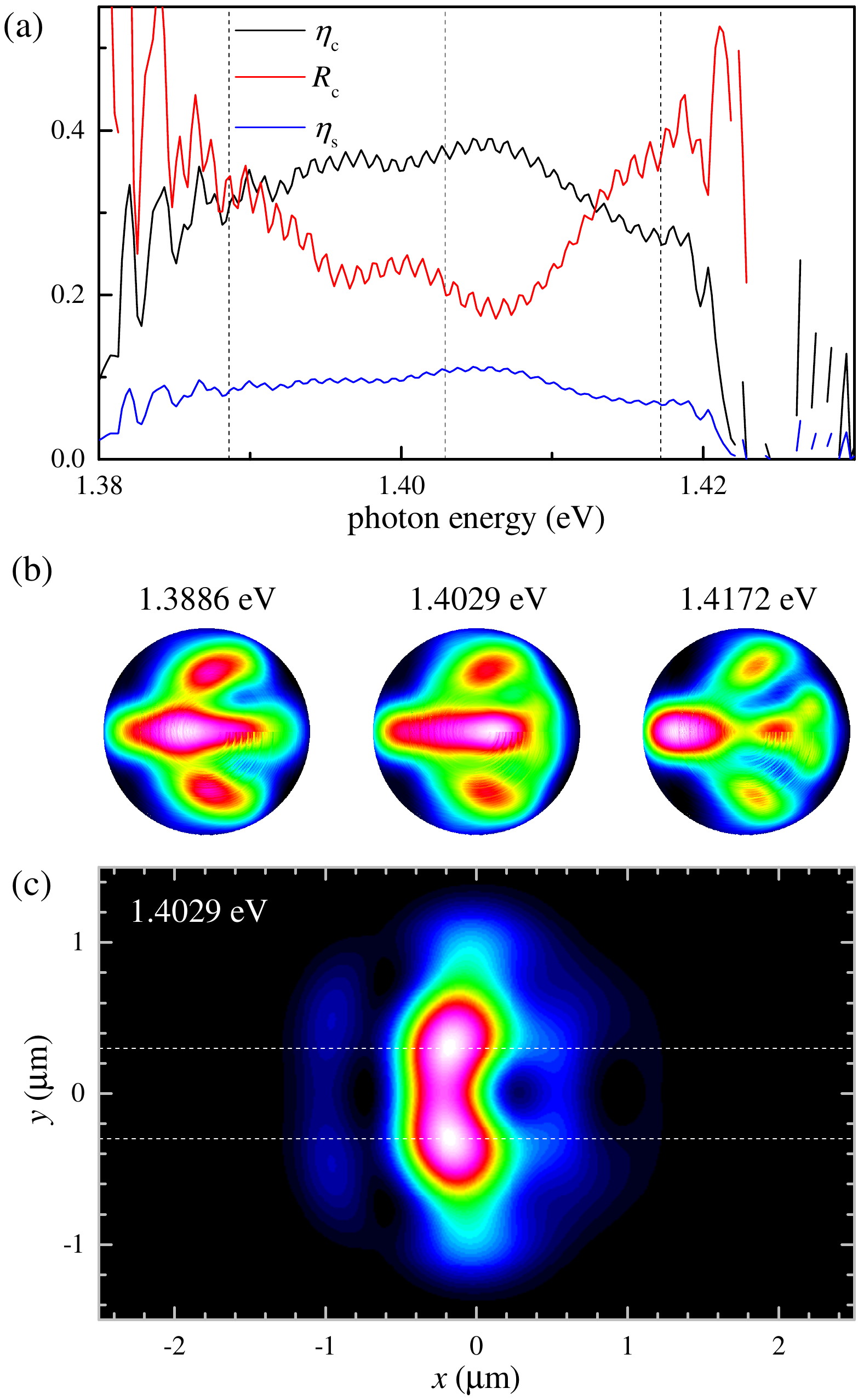}
	\caption{As \Fig{fig:grating}, but for the even (higher order) WG mode.}
	\label{fig:even_grating}
\end{figure}

In order to simulate the efficiency of coupling into the spectrometer, the far field of the couplers was extracted from the simulation. The MEEP near-field to far-field transform is used to calculate the full 6-components electromagnetic field vector at points homogeneously sampling a hemisphere of radius $R=10^6 a$, much larger than any simulation feature. The resulting far-field intensity distribution in $k_x,k_y$ is indicated in \Fig{fig:grating}b for three frequencies. The far-field was then limited to the objective collection range, and transformed back into the near field. In detail, for each far-field point, having $x,y,z$ coordinates on the hemisphere, we calculate the in-plane wavevector $\bk =k_0[x,y]/R$. We then rotate the 6-component electromagnetic field at that point from the radial to the $z$ propagation direction, resulting in $E_z = H_z = 0$, and simulating the transformation of the emitted field by the aplanatic objective into its back focal plane. The imaging of the coupler onto the spectrometer slit is then calculated by multiplying the field at each point with a 2D plane wave $\exp(i\bk\br)$, and summing the vector-field plane waves from every far-field point within the microscope NA ($|\bk|<0.85 k_0$). This procedure provides the near-field in a small angle approximation, valid for the small NA of about 0.03 of the image at the spectrometer slit, which is magnified by a factor of 31.3. This near field (see \Fig{fig:grating}c) is then transmitted through the spectrometer slit (dashed lines), providing a near-field collection efficiency. The coupler efficiency, and the near-field collection efficiency, are multiplied to produce the spectrometer efficiency \etas\ (see \Fig{fig:grating}a), which averages to 14.4\% in the simulated range. This is the efficiency with which photons emitted by a QD into the PCWG mode will enter the spectrometer, neglecting propagation loss in the WG. Losses occurring for all detected light, such as reflection loss of the optics, and detector quantum efficiencies, are not considered here as they are not influencing the measured \bfac\ \bfm.
These are estimated to provide an additional factor of around 30\% in the setup used, resulting in an estimated overall detection efficiency of QD emission around 4\%. 

The analysis above used the sum of the intensities of the two polarization components of the propagating field. To take into account the polarization dependence of the detection efficiency, due to the diffraction grating of the spectrometer, we separated the intensity transmitted through the spectrometer slit into $x$ and $y$ polarized components, and find 18\% of the power in $x$ and 82\% in $y$ polarization, at $E=1.3934$\,eV.

We performed an equivalent analysis for the even mode, as shown in \Fig{fig:even_grating}. This mode shows two minima and one maximum within the PC bandgap. Close to these extrema, the mode group velocity tends to zero, thus producing a slow light regime, which gives rise to difficulties in the numerical treatment, leading to unreliable results above 1.42\,eV and below 1.385\,eV in our simulations. In the remaining range, the calculated coupler efficiency is slightly lower and the reflectivity is significantly higher than for the odd mode.
The far-field pattern (see \Fig{fig:even_grating}b) is more structured, and the near-field (see \Fig{fig:even_grating}c) is wider than for the odd mode. Overall this results in a collection efficiency \etas\ of the higher order mode of 10\% at $E$=1.403\,eV, of which 73\% is $x$ polarized. 

To determine the corrected $\tilde{\bfs}$ for \Eq{Eq:ErrorPlotTerms}, we have used a collection efficiency of 14\% for the fundamental mode, calculated as the average over the simulated range, and 10\% for the higher order mode.

\subsection{Beta factor and directionality}
\label{subsec:bfaccirc}

\begin{figure}
	\centering	
	\includegraphics[width=\columnwidth]{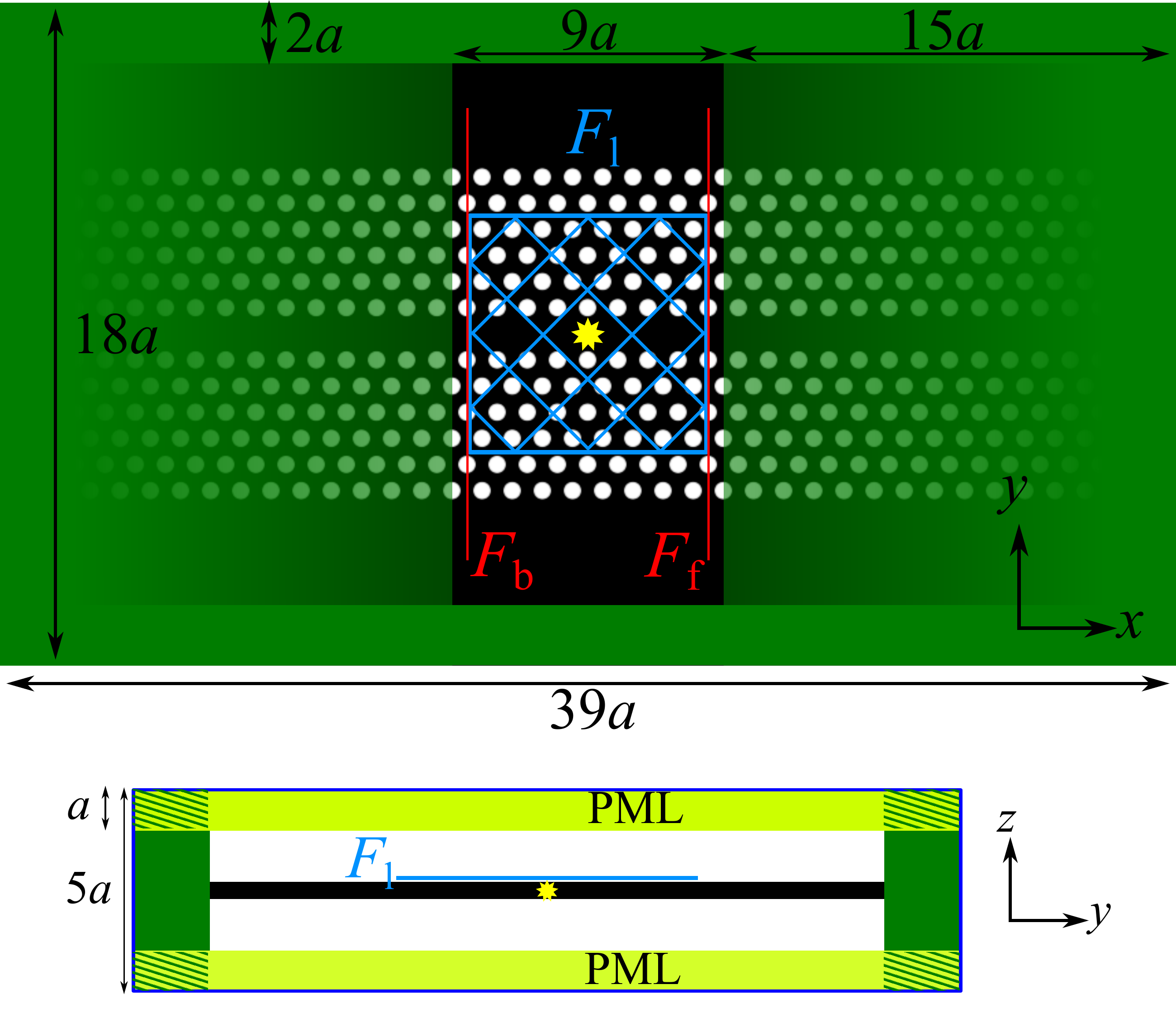}
	\caption{Sketch of the simulation volume used to calculate the \bfac\ and directionality. Colours and symbols have the same meanings as in \Fig{fig:grating_sim_map}.}
	\label{fig:Beta_fact_simulation}
\end{figure}

The \bfac\ and directionality were calculated using further FDTD simulations. We use of a simulation volume of size $39a$ in $x$, $18a$ in $y$, and $5a$ in $z$ direction, as sketched in \Fig{fig:Beta_fact_simulation}. A circular dipole source, $\mathbf{d} = [i, 1]$, is placed in the central unit cell at $z=0$,  at a given position in $x,y$. The source has the same time-dependence as used for the coupler simulation in \App{Sec:Grating Couplers}. The \bfac\ and directionality of the source is calculated from the power the source radiates into the different channels. To obtain accurate results, we have to suppress reflections from the boundaries of the simulation volume. Similar to \App{Sec:Grating Couplers}, we use adiabatic absorbers of $15a$ thickness covering the $\pm x$ boundaries of the simulation \cite{OskooiOE08}, leaving the central 9 unit cells unperturbed. In $y$ direction, the photonic band gap results in a strong reduction of the field, and the boundaries are in the region of unstructured slab, so that absorbers with thicknesses of $2a$ were found to suffice.
In $z$-direction, extending in free-space, PMLs of thickness $a$ were used. In the $\pm x$ directions the absorption profile is quartic while in the others it is quadratic. The mirror symmetry at $z=0$ is exploited in the calculation.

Flux planes in the WG forward and backward direction record the frequency-resolved transmitted powers $\Ff$ and $\Fb$, respectively. They are placed $4a$ away from the source in $x$, and are $15a$ wide in $y$, covering the entire photonic cladding but not the $\pm y$ absorbers.  In $z$ they extend $0.05a$ beyond either side of the slab, to contain the mode in the WG and most of its evanescent tail, while limiting the overlap with non-guided light. A third flux plane records the power radiated out of the slab ($\Fl$) in $+z$ direction, which is doubled to account for the symmetry. These three quantities determine the \bfac\ and the directionality by:

\be \bfs = \frac{\Ff+\Fb}{\Ff+\Fb+\Fl}\,, \quad \Ds = \frac{\Ff-\Fb}{\Ff+\Fb}. \ee

Note that $\bfs$ refers to the total fraction of the QD emission into all guided modes and at each frequency we consider there can be up to 6 such modes.

\begin{figure}
	\centering	
	\includegraphics[width=\columnwidth]{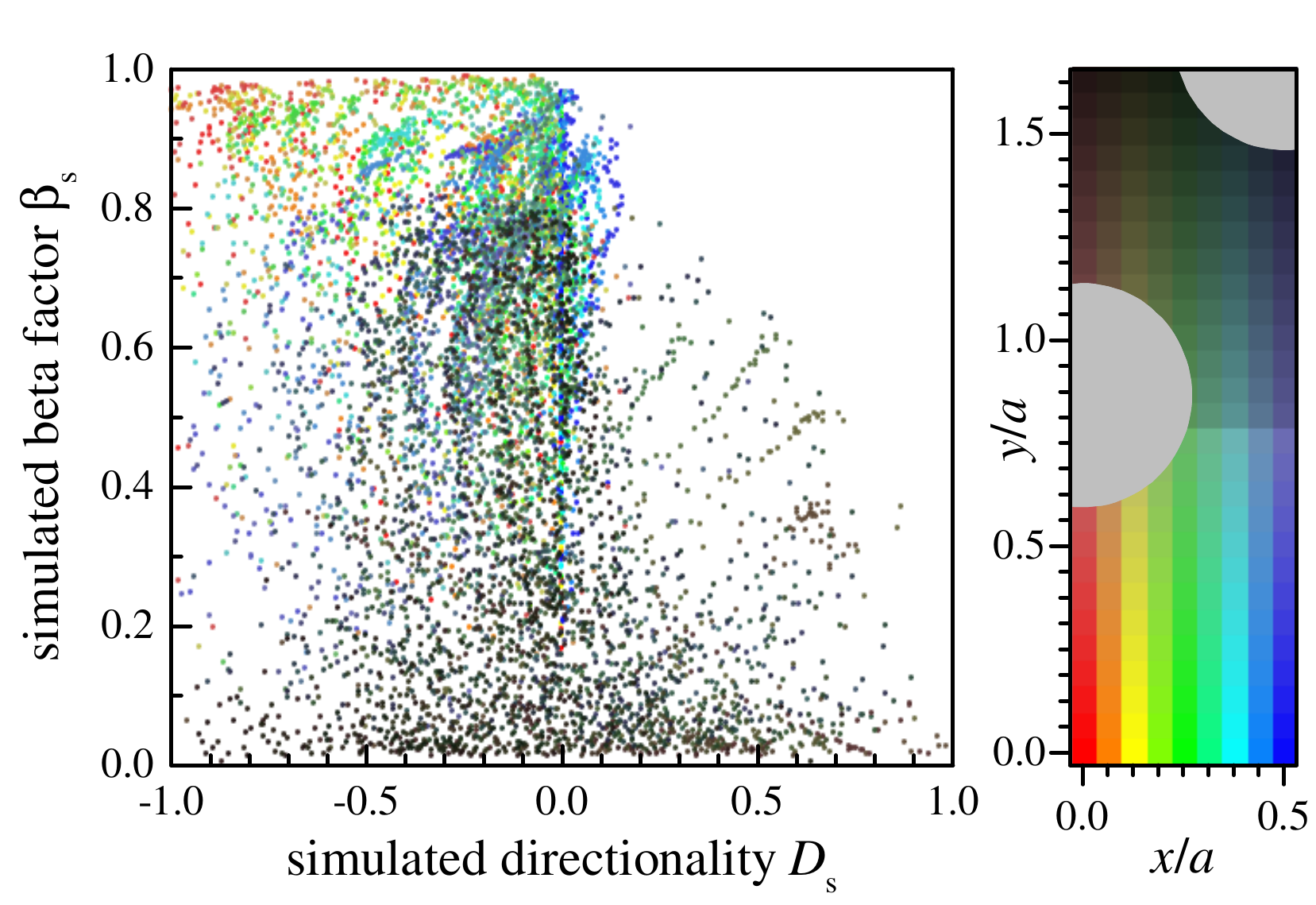}
	\caption{Simulated \bfac\ \bfs\ versus simulated directionality \Ds\ for 100 equidistant energies from 1.3457\,eV to 1.4411\,eV. The colour of the data points corresponds to their locations as shown in the right panel. Positions inside the holes (light grey) are not considered.}
	\label{fig:circ_beta}
\end{figure}

In a single mode waveguide hosting a circular dipole source, $D$ at each location and frequency is equal to the third Stokes parameter $S_3$ of the mode field. In multimode waveguides $D$ is given by an average of the mode $S_3$'s weighted by their densities of states \cite{LangPTRSA16}. For comparison with experiment $D$ can be converted into a circularity value by:

\be
C = \frac{1}{2}\log\left( \frac{1 + D}{1 - D}\right).
\ee

In each simulation these quantities are extracted as a function of frequency for a given source location. To probe the spatial dependence of the \bfac\ and circularities, simulations with the source dipole at different locations were performed. As the dipole source is moved in the $x$ direction it is moved closer to one waveguide flux plane and further from the other. For modes with propagation loss this will impact the measured directionality as the light travels further in one direction than the other. To compensate for this, all calculations were run twice, once with the dipole at $+x$ and again at $-x$. The lattice symmetry ensures that the directionality must be identical at these two locations. Averaging the powers of the two simulations accounts for the effect of being closer to one flux plane or the other, removing the impact of losses on the directionality, as long as they are small over a unit cell.  The effect of the losses on the \bfac\ remain, due to the propagation loss between the source location and the forwards/backwards waveguide flux planes. We can see in \Fig{fig:loss} that the propagation loss over this distance of about 1\,\textmu m is negligible ($<3$\%) for most of the modes we consider. The only exception is the lossy branch of the even mode, which experiences significant attenuation over this distance. The energy lost by this mode is radiated out of the slab and collected by the flux plane measuring the free space emission. The mode is so lossy that it is part of the free space emission, and this is the case also in the measurements. 

The \bfac s inferred from these calculations are plotted in \Fig{fig:circ_beta} versus the directionality. Each data point corresponds to a location in the WG unit cell and an emission energy. The locations are colour coded according to the map shown. The hue varies along the $x$ axis and the saturation along $y$. Note that the breaking of symmetry in $D$ originates from the display of only half the unit cell, with the other half being mirror symmetric, mapping $x$ to $-x$ and $D$ to $-D$.

To assess the sensitivity of these results on the choice of the etch parameter $d$, simulations were performed at 5 random locations with $d=7$\,nm instead of $8$\,nm. After accounting for an overall frequency shift of $0.0014 (c/a)$ (6.7\,meV) the mean absolute change of $\beta$ was 0.024, and the change in directionality $D$ was 0.014. These small changes reflects the fact that the main effect of altering the etch parameter is a shift in frequency while hardly changing the mode profiles.

As discussed in \App{Sec:Grating Couplers}, these simulations will be inaccurate close to the extrema of the even mode, and these energies are not shown in \Fig{fig:circ_beta}.

\begin{figure}
	\centering
	\includegraphics[width=\columnwidth]{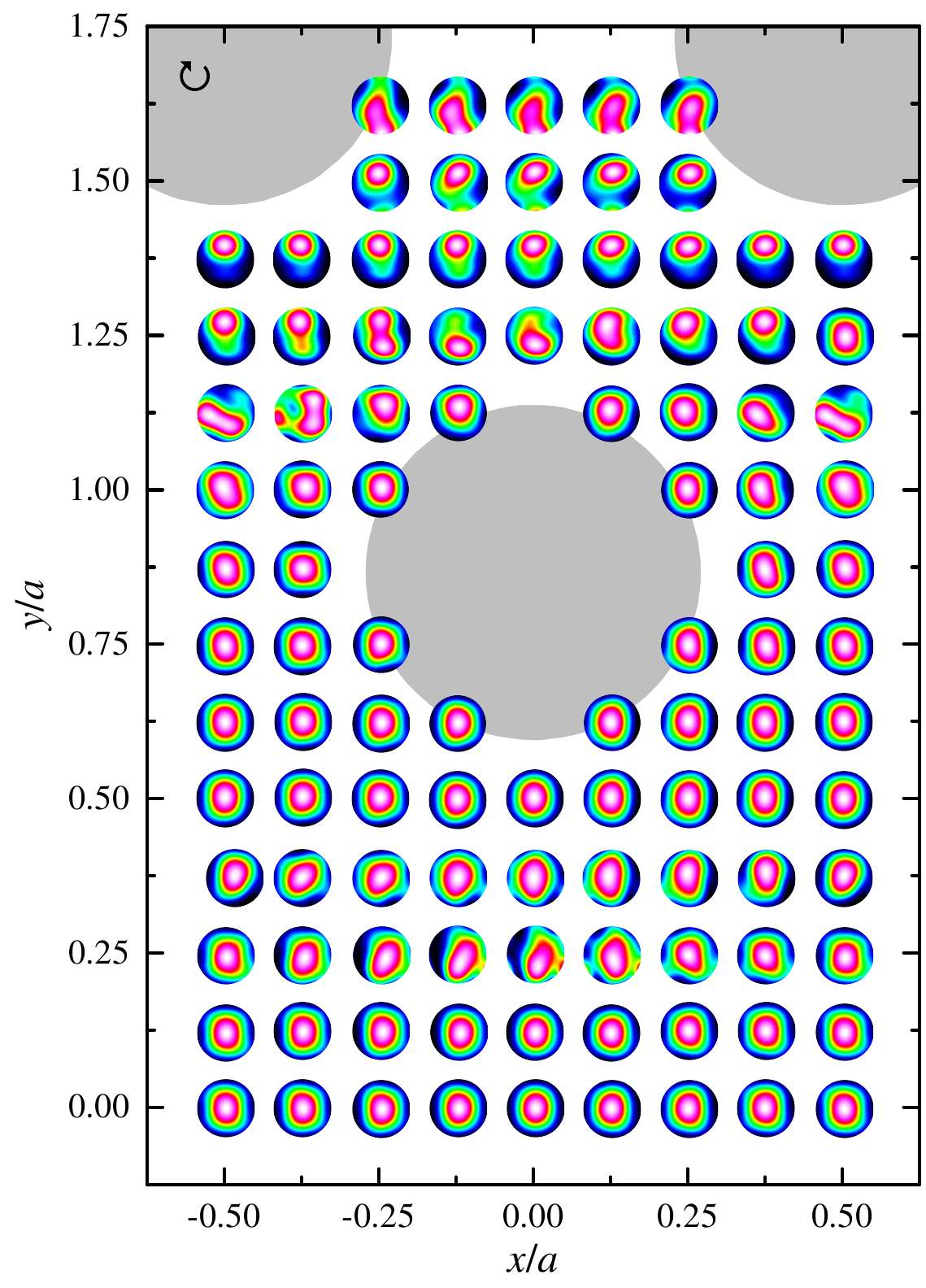}
	\caption{Far-field profiles of the free space emission of a right-handed ($\circlearrowright$) circularly polarised dipole source at $z=0$ with emission energy $E$=1.3934\,eV, as a function of its position $x,y$ within the PCWG unit cell. The circular plots are in $\bk$ space having radius of $k_0$.}
	\label{fig:FSfarfield}
\end{figure}
\begin{figure*}
	\centering
	\includegraphics[width=\textwidth]{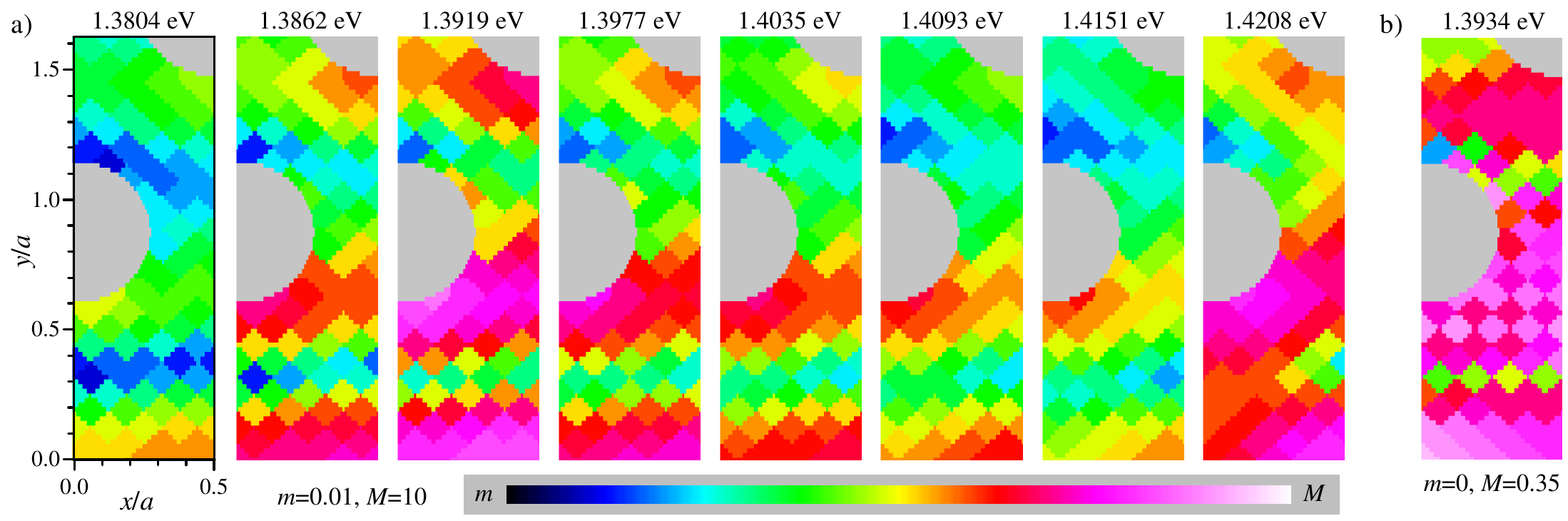}
	\caption{a) Emission power for a circular dipole source into non-guided modes, relative to the emission power in bulk GaAs,  as a function of source position, for different emission energies as given. Logarithmic colour scale from $m$ to $M$ as given. b) collection efficiency of the emission of a right-handed ($\circlearrowright$) circularly polarised dipole source with emission energy $E$=1.3934\,eV, as a function of the source position, including the objective NA and the input slit of the spectrometer. Linear colour scale from $m$ to $M$ as given.}
	\label{fig:FSCollection}
\end{figure*}
\subsection{Free-space emission}
\label{subsec:freespace}
The far field intensity distributions in $\bk$ space of the free space emission of circular dipoles at various locations in the PCWG unit cell, taken from the simulations discussed in \App{subsec:bfaccirc}, are shown in \Fig{fig:FSfarfield} for $E=1.3934$\,eV. They show a significant variation with source location. Note that the missing mirror symmetry about $x=0$ arises from the  circularly polarized dipole excitation which lacks this symmetry. Using these profiles, we determine the corresponding free space collection efficiency into the spectrometer, as shown in \Fig{fig:FSCollection}b, considering the objective NA and the spectrometer input slit as described in \Sec{Sec:Grating Couplers}. We can observe that the regions of low values correspond to highly asymmetric far-field profiles in \Fig{fig:FSfarfield}. The average efficiency is about 29\%, and varying from 18\% to 33\% with the location of the QD. We can consider these collection efficiencies to improve the accuracy of the \bfac s determined in the experiment. We note that free space emission is collected with up to a factor of 2 higher efficiency than the WG mode emission, suggesting that the real \bfac s are up to a factor of two closer to unity than reported in the main text. 

To investigate the validity of the assumption of free-space emission strength independent of position, used in previous works reporting on the \bfac, we show in \Fig{fig:FSCollection}a the free space emission power as function of position, relative to bulk GaAs, for 8 energies covering the simulation range. We observe variations over two orders of magnitude, from 0.05 to 5, implying that the assumption is not justified. This points towards large systematic errors in the \bfac s retrieved in previous works \cite{Lund-HansenPRL08,ThyrrestrupAPL10,DewhurstAPL10,LauchtPRX12,HoangAPL12,ArcariPRL14}. Large spatial variations in the free space emission are also indicated by the simulations in \cite{JavadiJOSAB18}, which also predicts horizontal bands of reduced free space coupling as seen in \Fig{fig:FSCollection}.

For $E$=1.4208\,eV, we see large increases in the loss rate. This is attributed to the branch of the even mode in the light cone, as the energy is close to the frequency maximum in the even mode, where its group velocity tends to zero.  For a low group velocity the emission is enhanced by the slow light, and the propagation length is reduced, resulting in an emission into free space close to the emitter. Indeed, the pattern resembles the electric field intensity pattern of the even (higher order) seen in \Fig{fig:FieldProfile}.

\subsection{Mode Separation}\label{Sec:ModeSeparation}

The FDTD calculations provide the \bfac\ and directionality. The values observed in the experiment will be slightly modified from these values by the different collection efficiencies of coupler and free space emission. As the collection efficiencies of the even and odd mode are different, adjusting for these efficiencies requires separation of the power between the two modes. To achieve this separation, the electrical field profiles $\mathbf{E}(\mathbf{r}, \omega)$) and group indices $n_{\rm g}(\omega)$ extracted from the loss FDTD simulations (see \App{loss}) are used. The group indices are calculated using the spatially integrated Poynting vector \cite{ChenPRB10}. These are used to calculate the density of states (DOS) into the forward and backward directions at each spatial location using the Purcell factor expressions presented in \cite{MangaRaoPRB07}, adapted to separate the intensity in a given direction. As we are only concerned with ratios between modes at a common frequency, these expressions can be simplified to

\be
\begin{split}
	\text{DOS}_{\text{Forwards}} \propto |\mathbf{E}(\mathbf{r}, \omega)|^2 n_g(\omega) (S_3 + 1)/2\\
	\text{DOS}_{\text{Backwards}} \propto |\mathbf{E}(\mathbf{r}, \omega)|^2 n_g(\omega) (1 - S_3)/2.
\end{split}
\ee

Using the fields associated with each mode, respectively, allows us to calculate the fraction of the power emitted into the forward even mode, for a given location and frequency. We neglect the branch of the even mode in the light cone, as the power in it is lost while propagating to the coupler. Using the resulting fractions the collection efficiencies of the couplers can be corrected while accounting for the different efficiencies for the two modes. 

\section{Correction of simulations}
In this section we describe how the raw data presented in \Sec{subsec:bfaccirc} have been corrected in order to be comparable to the experimental data. The forward $f$ and the backward $b$ propagating fluxes are firstly separated into the fundamental and higher order modes projections as calculated in \Sec{Sec:ModeSeparation}. The fluxes of each mode are then multiplied by the corresponding collection efficiencies due to the couplers, the NA of the MO and the input slit of the imaging spectrometer. We recall that the collection efficiencies are averaged over the spectral range covered by the simulations. The resulting fluxes from each mode are then separated into transverse and longitudinal polarisation components using the analysis presented in \Sec{Sec:Grating Couplers}. Finally, we include the polarisation dependent efficiency $T$ and $L$ of the spectrometer grating for transverse and longitudinal polarisations respectively. The polarisation resolved fluxes are multiplied by the coefficients $T$ and $L$, and the resulting fluxes are then summed up, separately for each mode, to obtain the final corrected fluxes. $T$ and $L$ are determined by measuring the grating polarisation efficiency ratio $R=T/L$, with the normalisation $T+L=2$, such that for polarization independent detection we have  $T=L=1$. 

For the free space emission we first multiply the out-of-plane flux by the position dependent collection efficiency, as described in \Sec{subsec:freespace}. In this case the collection efficiency is calculated at $E=1.3934$\,eV, which is approximately the central energy of the measured QDs. We calculated the energy dependence at one location and we found it to be negligible. In order to include the polarisation efficiency of the grating spectrometer, we consider the free space emission to be circularly polarised. 
%
%

\end{document}